\documentclass[5p,twocolumn]{elsarticle}
\usepackage{hyperref}
\usepackage{amssymb}
\usepackage{subcaption}
\usepackage{natbib}

\journal{Icarus}

\biboptions{authoryear}

\begin{document}

\begin{frontmatter}

\title{On the sodium enhancement in spectra of slow meteors and the origin of Na-rich meteoroids}

\author[label1]{Pavol Matlovi\v{c}}\ead{matlovic@fmph.uniba.sk}
\author[label1]{Juraj T\'{o}th}
\author[label1]{Leonard Korno\v{s}}
\author[label2]{Stefan Loehle}
\address[label1]{Faculty of Mathematics, Physics and Informatics, Comenius University, Bratislava, Slovakia}
\address[label2]{High Enthalpy Flow Diagnostics Group, Institut f\"ur Raumfahrtsysteme, Universit\"at Stuttgart, Pfaffenwaldring 29, D-70569 Stuttgart, Germany}

\begin{abstract}
The detected Na/Mg ratio in a sample of 17 Na-enhanced and Na-rich meteors is investigated based on obtained spectral, orbital and structural data. We utilize the meteor observations of the AMOS network obtained within a survey of medium-sized meteoroids supplemented by higher-resolution spectra observed on the Canary Islands. Ground-based meteor observations are then compared with high-resolution Echelle spectra of simulated ablation of known meteorite samples in a high-enthalpy plasma wind tunnel. It was found that most Na-enhanced and Na-rich spectra can be explained by the effect of low meteor speed related to low ablation temperatures and generally do not reflect real meteoroid composition. Spectra obtained by the laboratory experiment simulating low meteor speeds show corresponding Na-rich profiles irrespectively of the meteorite composition. We estimate that for an H-type ordinary chondrite with speed of $\approx$ 10 km\,s\textsuperscript{-1}, the Na line intensity is increased by a factor of 40 to 95. The dynamical analysis has revealed that all Na-rich meteors originated on Apollo-type orbits and exhibit consistent chondritic material strengths.  For more clarity in the classification of Na-enhanced and Na-rich meteoroids, we propose new speed-dependent boundaries between the spectral classes. Real compositional Na enhancement was confirmed in five cometary meteoroids: two Perseids, an $\alpha$-Capricornid, $\nu$-Draconid and a sporadic. The two Na-enhanced Perseids were linked with increased material strength suggesting that the detected increase of volatile content has implications for the meteoroid structure.
\end{abstract}

\begin{keyword}
Meteors\sep Meteorites \sep Asteroids \sep Comets \sep Spectroscopy
\MSC[2010] 85-02\sep  85A04
\end{keyword}

\end{frontmatter}


\section{Introduction}

Even though sodium does not represent a major compositional component in interplanetary bodies, it is of high cosmochemical interest. It is assumed that volatile elements such as sodium were depleted from the early inner solar protolanetary disk due to intense solar radiation, and the later formed rocky bodies probably suffered differentiation that further eroded volatiles \citep{1997Icar..130....1E}. Therefore, higher sodium abundances are expected for cometary bodies formed in the outer regions of the solar system \citep[see the discussion in][]{2004MNRAS.348..802T, 2007AdSpR..39..517T}. It has been shown that Na content in meteoroids can be used to infer details about the structure and thermal history of meteoroids \citep{2005Icar..174...15B, 2019A&A...621A..68V}. Similarly to cometary studies, we can take advantage of the fact that sodium is a very volatile element with low excitation and high luminous efficiency. As pointed out by \citet{2004MNRAS.348..802T}, the spectra of large meteoroids observed in the atmosphere show overabundance of sodium compared to the measured composition of chondritic meteorites on Earth's surface and interplanetary dust particles studied by space probes, as a result of the volatile loss during the atmospheric interaction.

Studies of the temporal evolution of meteor spectra pointed out the specific behavior of the Na line. The early release of sodium in Draconid meteors was first noted by \citet{1972JRASC..66..201M} and further analyzed by \citet{2007A&A...473..661B} and \citet{2014EM&P..113...15B}. Alternatively, in spectra of Leonid meteors \citep{1999M&PS...34..987B}, Na line appeared early, but disappears abruptly. In other cases, including our own experiences \citep{2017P&SS..143..104M, 2019A&A...629A..71M}, sodium is often observed to be the first to appear and the last to disappear. The pattern of sodium ablation is even more complex if we consider differences between sporadic meteors and shower meteors. Apparently, the early release of sodium is not universal and must depend on meteoroid structure.
   
It is assumed that sodium may be released preferentially at the beginning of the luminous trajectory if the body is disrupted early \citep{2007A&A...473..661B}. The quick evaporation of sodium from the entire volume observed in Leonid meteoroids can be explained by the dust-ball model, which considers initial disruption of mm-sized meteoroids. In the dust-ball model  \citep{1975MNRAS.173..339H}, meteoroids are composed of grains held together by volatile low boiling point material. Sodium is often interpreted as part of this "glue", which after reaching its melting point, releases meteoroid grains. However, it was shown that the presence of this glue is not necessary to explain the early release of Na. Na can also be part of the constituent grains. Differential ablation is expected for small dust particles \citep{2007A&A...473..661B} and theoretical computations \citep{1998JGR...10310899M} concluded that Na can evaporate completely from the whole volume before Mg starts to ablate. Furthermore, the study of \citet{2019A&A...621A..68V} has shown that larger Na content is often associated with meteoroids composed of larger grains and probably higher porosity.

The first survey of meteor spectra \citep{2005Icar..174...15B} has indicated that while many meteoroids have properties similar to primitive chondrites, numerous atypical spectral classes can be found. These deviations often relate the detected increased or depleted Na content. \citet{2005Icar..174...15B} have reported three populations of Na-free meteoroids. The source of sodium depletion was related to the processes of thermal desorption in meteoroids on orbits with small perihelion distance and cosmic ray irradiation in meteoroids on Halley-type orbits. Natural Na deficiency is also observed in spectra of iron meteoroids. Similar meteoroid types were later detected among larger cm-dm meteoroids \citep{2019A&A...629A..71M}. Here the Na-poor and Na-free meteoroids comprised smaller part of the sample and overall Na enhancement was detected compared to mm-sized bodies, showing that volatiles are better preserved in larger meteoroids.   
   
Besides the Na-poor and Na-free meteors, the surveys of \citet{2005Icar..174...15B} and later \citet{2019A&A...621A..68V} have revealed the presence of Na-enhanced and Na-rich meteors among mm-sized bodies. These spectral types were more commonly found among larger meteoroids studied by \citet{2019A&A...629A..71M}. While the Na depletion in meteor spectra is directly linked to meteoroid composition affected by space weathering, the source the Na enrichment has not been explained. The situation is complicated by the fact that Na/Mg intensity ratio commonly used to distinguish Na enriched meteors depends on meteor speed, as a result of the low excitation and ionization potential of Na \citep{2005Icar..174...15B}. This dependency is only empirical and not well defined particularly for very slow meteors ($<$ 15 km\,s\textsuperscript{-1}). Furthermore, only limited orbital and physical data for Na-rich meteors were obtained in previous studies. 

Here, we aim to evaluate the real Na content and influence of meteor speed on the detected Na/Mg ratio based on a larger sample of Na-enhanced and Na-rich spectra observed by multiple stations. The obtained wide set of parameters allows us to analyze the dynamical source and material strengths of meteoroids with Na-enhanced and Na-rich spectra. Furthermore, the atmospheric observations are confronted with simulated ablation of meteorites, allowing us to study the emission spectra of samples with known composition at comparable conditions.

First, in Section \ref{candidates}, we describe the properties of candidate meteors initially classified as Na-enhanced and Na-rich. The justification of this classification with respect to the speed, dynamical origin and physical properties of these bodies as well as the meteorite emission spectra obtained during a laboratory experiment is investigated in Section \ref{SpeedDep}. Based on the results of this analysis, the proposed new boundaries for the spectral classification and the confirmed Na-enhanced and Na-rich meteoroids are described in Section \ref{Reclass}.
 
\begin{figure}
\centerline{\includegraphics[width=8cm,angle=0]{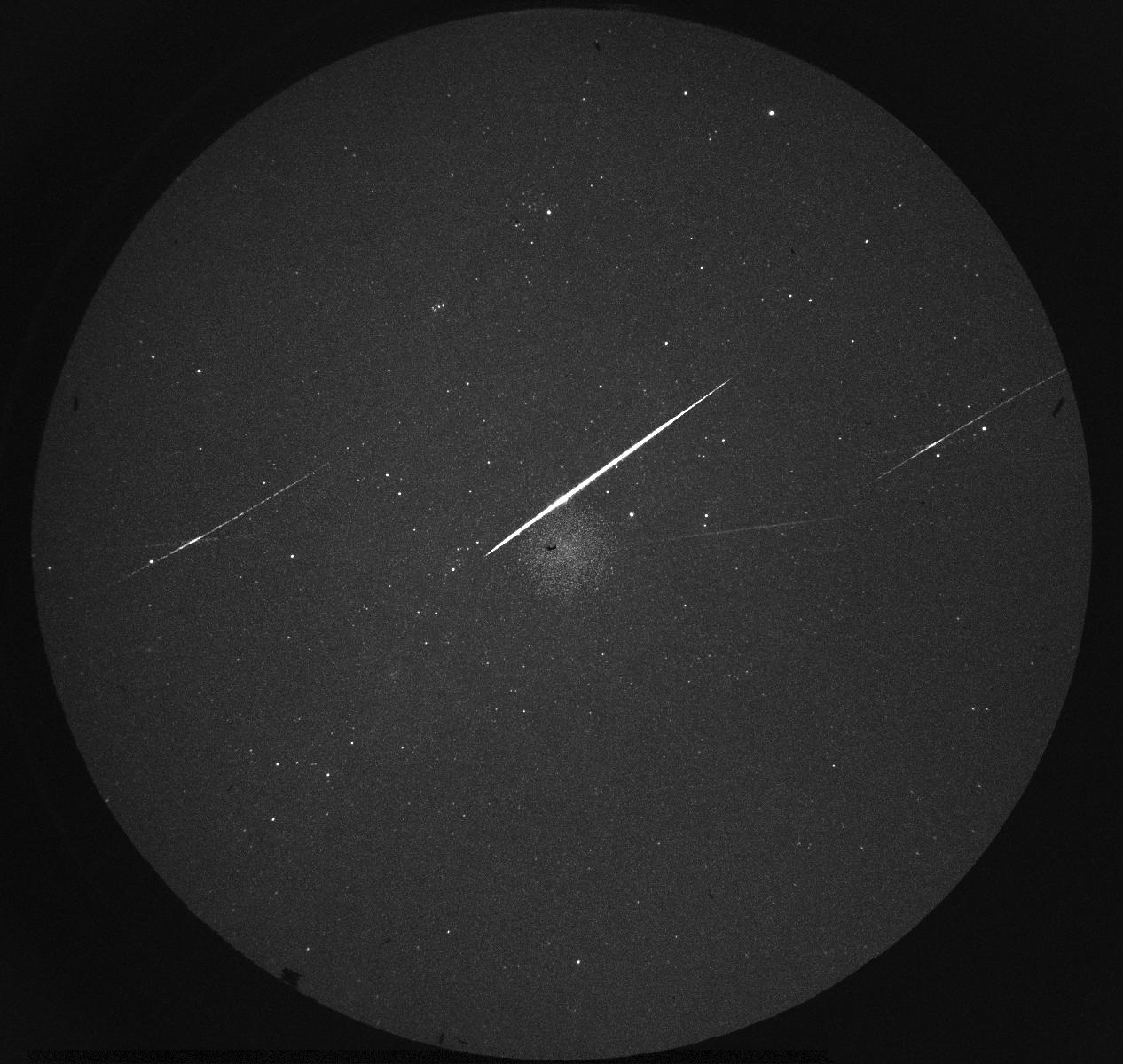}} \caption[f1]{Na-rich meteor observed by the AMOS-Spec system on February 13, 2015. The meteor (0-order) is captured in the middle of the FOV along with the +1 and -1 order of its spectrum consisting solely of the Na I line.} 
\label{NaRFOV}
\end{figure}

\section{Methods and observations}

The presented analysis is primarily based on 17 multi-station Na-enhanced and Na-rich meteor spectra observed within the global All-Sky Meteor Orbit System (AMOS) network. The studied Na-enhanced and Na-rich meteors were identified within the survey presented in \citet{2019A&A...629A..71M}. Additional Na-rich meteors were obtained from newer observations of the AMOS-Spec system in Modra, Slovakia and higher-resolution observations by AMOS-Spec-HR systems at the Teide observatory on Tenerife and Roche de los Muchachos observatory on La Palma, Canary Islands. More details about the global AMOS network are available in \citet{2015P&SS..118..102T} and \citet{TothIMC17}.

\subsection{Meteor observations and instrumentation}

The AMOS-Spec is a semi-automatic remotely controlled video system for systematic meteor spectra observations. The system setup yields a wide 100$^{\circ}$ circular field of view (FOV) with a resolution of 1600x1200 px and frame rate of 12 fps. Diffraction of the incoming light is provided by a holographic grating with 1000 grooves/mm placed above the lens. The resulting dispersion varies due to the geometry of the all-sky lens with an average value of 1.3 nm/px. This translates to spectral resolution of approximately R $\approx$ 165. The limiting magnitude of the AMOS-Spec for meteors (zero order) is around +4 mag, while the limiting magnitude for spectral event (first order) is -1 mag. The system covers the whole visual spectrum range of 370 –- 900 nm with sensitivity peak at 470 nm. Example of a Na-rich meteor observed by the AMOS-Spec system is on Fig. \ref{NaRFOV}. More information about the AMOS-Spec system can be found in \citet{2019A&A...629A..71M}.

\begin{figure}[t]
\centerline{\includegraphics[width=8cm,angle=0]{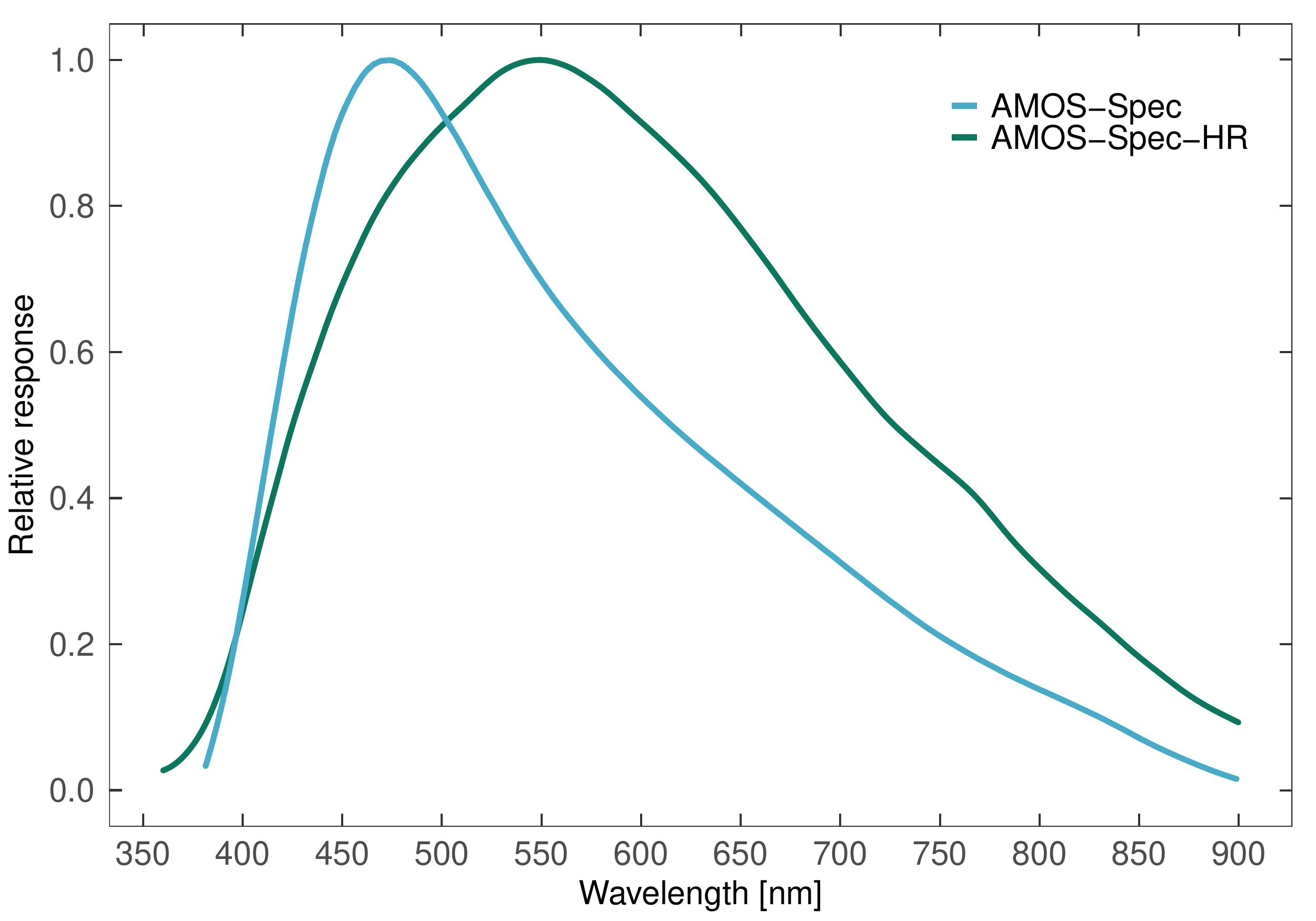}} \caption[f1]{Spectral response curve of the lower-resolution system AMOS-Spec and the higher-resolution AMOS-Spec-HR.} 
\label{Sens}
\end{figure}

Since November 2016, the AMOS network also consists of higher-resolution spectrographs AMOS-Spec-HR which accompany the AMOS stations in Canary Islands, Chile and Hawaii. This system is based on 2048x1536 px Point-Grey camera, 6 mm f/3.5 lens, and 1000 grooves/mm holographic grating, resulting in 60x45$^{\circ}$ rectangular FOV and dispersion of 0.5 nm/px. The spectral resolution of the AMOS-Spec-HR is R $\approx$ 500. The limiting magnitude of the system is around -1.5 mag for spectral events, and obtained recordings have frame rate of 15 fps. The spectral response curves of AMOS-Spec and AMOS-Spec-HR systems is on Fig. \ref{Sens}.

Each analyzed spectrum was manually scanned in individual frames, calibrated and fitted according to the procedure described in \citet{2019A&A...629A..71M}. After the reduction of the continuum radiation and atmospheric emission lines and bands, which are generally very faint in spectra of slow meteors, all present meteor emission lines were identified and measured. Intensity ratios the main spectral multiplets of Na I - 1, Mg I - 2 and Fe I - 15 were determined from the fitted spectrum and used to identify Na-enhanced and Na-rich meteors, in accordance with the spectral classification method introduced by \citet{2005Icar..174...15B}.

All of the meteoroids presented in this work were observed by multiple stations of the AMOS network, providing orbital and trajectory parameters. The standard AMOS system consists of four major components: a fish-eye lens, image intensifier, projection lens and a digital video camera. The resulting FOV of AMOS systems is 180$^{\circ}$ x 140 $^{\circ}$ with the image resolution of 1280 x 960 pixels (8.4 arcmin/pixel) and frame-rate 15 fps for the original systems in Slovakia, and with resolution of 1600 x 1200 pixels (6.75 arcmin/pixel) and frame-rate 20 fps for the updated version used within AMOS network on Canary Islands, Chile and Hawaii. Limiting magnitude for stars is about +5 mag for a single frame, the detection efficiency is lower for moving objects approx. +4 mag due to the trailing loss.

The \textit{UFOCapture} detection software \citep{2009JIMO...37...55S} was used during real-time observations to detect and record all of the studied meteors. Star and meteor coordinates from each frame of each event were measured using the original \textit{AMOS} software. Meteor photometry, astrometry and orbit determination was performed using the \textit{Meteor Trajectory} (\textit{MT}) software \citep{2015pimo.conf..101K,KornosIMC17} based on the procedure of \citet{1987BAICz..38..222C} and \citet{1995A&AS..112..173B}. The precision of the AMOS astrometry is on the order of 0.02 - 0.03$^{\circ}$, which translates to an accuracy of tens of meters for atmospheric meteor trajectory.

\begin{table}[]
\small\begin{center}\caption {Meteoroid material strength classification based on parameters $K_B$ and $P_E$ following \citet{1988BAICz..39..221C}.} 
\begin{tabular}{lcc}
\hline\hline\\
\multicolumn{1}{l}{Material type}& %
\multicolumn{1}{c}{$K_B$ type}& %
\multicolumn{1}{c}{$P_E$ type} \\
\hline\\
Fragile cometary / Draconid-type & D & IIIB \vspace{0.2cm} \\
Regular cometary  & C & IIIA \vspace{0.2cm} \\
Dense cometary  & B & - \vspace{0.2cm} \\
Carbonaceous    & A & II \vspace{0.2cm} \\
Ordinary chondrite - asteroids & ast & I \vspace{0.1cm}\\
\hline
\end{tabular}
\label{KBPEclass}
\end{center}
\end{table}

The empirical parameters $K_B$ and $P_E$ \citep{1968SAOSR.279.....C, 1976JGR....81.6257C} were used to estimate the meteoroid material strength (Table \ref{KBPEclass}). It should be noted that the $K_B$ classification was originally developed for fainter meteors, while the $P_E$ classification focuses on fireballs. The resulting material characterization can thus show some discrepancies. As the meteors in our sample are generally of medium magnitudes (-2 to -7), both parameters were determined. 

\subsection{Meteorite ablation and laboratory instrumentation}

The presented meteor data from ground-based observations are compared with the results from the simulated ablation of meteorite samples in the high-enthalpy plasma wind tunnel at the Institute of Space Systems (IRS) of the University of Stuttgart. The meteorite samples were exposed to a high enthalpy plasma flow (mixture of N\textsubscript{2} and O\textsubscript{2}) in a vacuum chamber simulating the equivalent air friction of an atmospheric entry speed of about 10 km\,s\textsuperscript{-1} at 80 km altitude, resulting in the vaporization of the meteoritic material \citep{2017ApJ...837..112L}. These conditions recreate the atmospheric flight characteristic for the lower limit of meteor speeds.

Three meteorite samples (H and L chondrite and EL chondrite) were tested during our first experiment campaign and observed by higher resolution Echelle spectrograph \citep[also see][]{Loehle, 2017ApJ...837..112L}. The captured Echelle spectra cover the 250 -- 880 nm wavelength range with an average spectral resolution of $\sim$ 0.08 nm px\textsuperscript{-1} and a frequency of 10 spectra per second. The spectra for each sample were summed to produce representative overall spectral profile.

\section{The Na-enhanced and Na-rich spectral class} \label{candidates}

The spectral classification method introduced by \citet{2005Icar..174...15B} defines various spectral types of meteors based on the relative intensity ratios of emission multiplets Na I - 1, Mg I - 2 and Fe I - 15. Na-enhanced meteors are defined as having apparently higher Na line intensity than expected for chondritic composition (represented by normal type spectra) at given speed \citep{2005Icar..174...15B}. Other species, such as Mg I, Fe I are also present, but fainter in comparison. Moreover, spectra of Na-rich meteors are even more dominated by Na. The Na/Mg and Na/Fe ratios are significantly higher than chondritic, with Mg and Fe lines often almost undetectable at the noise level of the recording.

\begin{figure}
\centerline{\includegraphics[width=8.5cm,angle=0]{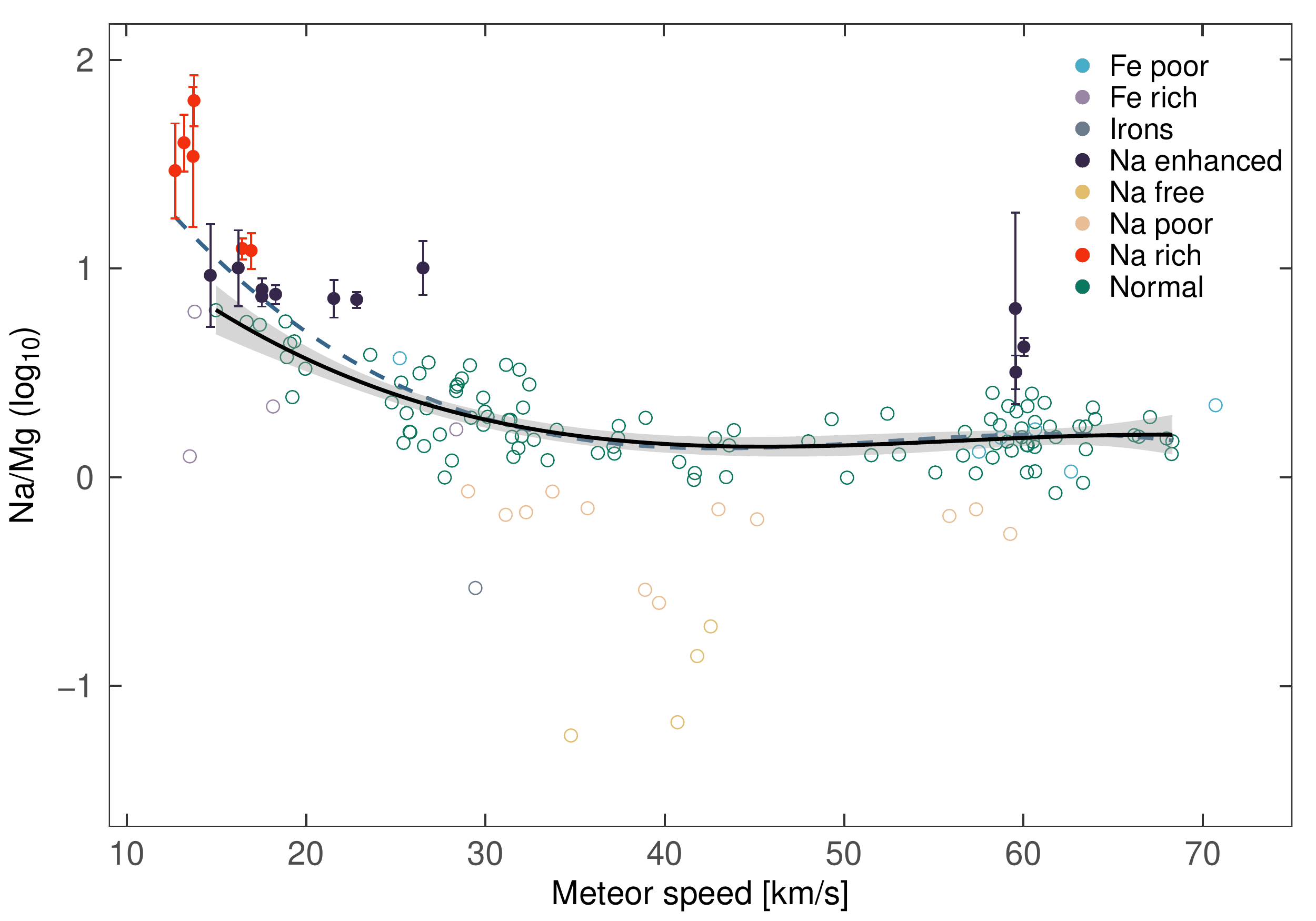}} \caption[f1]{The measured Na/Mg intensity ratio as a function of initial meteor speed based on the data from \citet{2019A&A...629A..71M}. The positions of the Na-rich and Na-enhanced meteors selected for analysis are highlighted. The solid line represents a cubic fit to the running average with standard error of the mean for meteors classified as normal. The dashed line represents a fit accounting for slow Na-rich and Na-enhanced meteors.} 
\label{speedNaErr}
\end{figure}

Meteor spectra are also affected by the physical conditions of the atmospheric flight, particularly by meteor speed, related to the temperature of the radiating plasma. \citet{2005Icar..174...15B} have shown that the Na/Mg intensity ratio is dependent on meteor speed below 40 km\,s\textsuperscript{-1} as a result of the low excitation and ionization potential of Na compared to Mg. We have found similar dependency for spectra of larger cm-sized meteoroids \citep{2019A&A...629A..71M}. The speed dependency on the Na/Mg intensity ratio and the selection of meteors for our analysis are displayed in Fig. \ref{speedNaErr}. Furthermore, meteoroid size also affects the detected Na content, as cm-dm meteoroids show overall increase of Na/Mg intensity compared to mm-sized particles \citep{2019A&A...629A..71M}, reflecting better preservation of volatiles in larger bodies. 

\subsection{Nomenclature}

Given the current disparity and unclear boundaries for the classification of Na-enhanced and Na-rich meteoroids, the sample analyzed here should be understood as a set of candidates for the Na-enhanced and Na-rich classes, as identified in our previous survey \citep{2019A&A...629A..71M}. Here we will denote such candidates (i.e. meteors having spectra with apparently strong/dominant Na line) as Na-enhanced/Na-rich \textit{meteors} or Na-enhanced/Na-rich \textit{spectra}. The cases where real composition with larger than normal abundance of Na is reported will be denoted as Na-enhanced/Na-rich \textit{meteoroids}.

\subsection{Na-enhanced meteors}

\begin{figure*}[t]
    \centering
    \begin{subfigure}[b]{0.5\textwidth}
      \centering
      \includegraphics[width=\textwidth]{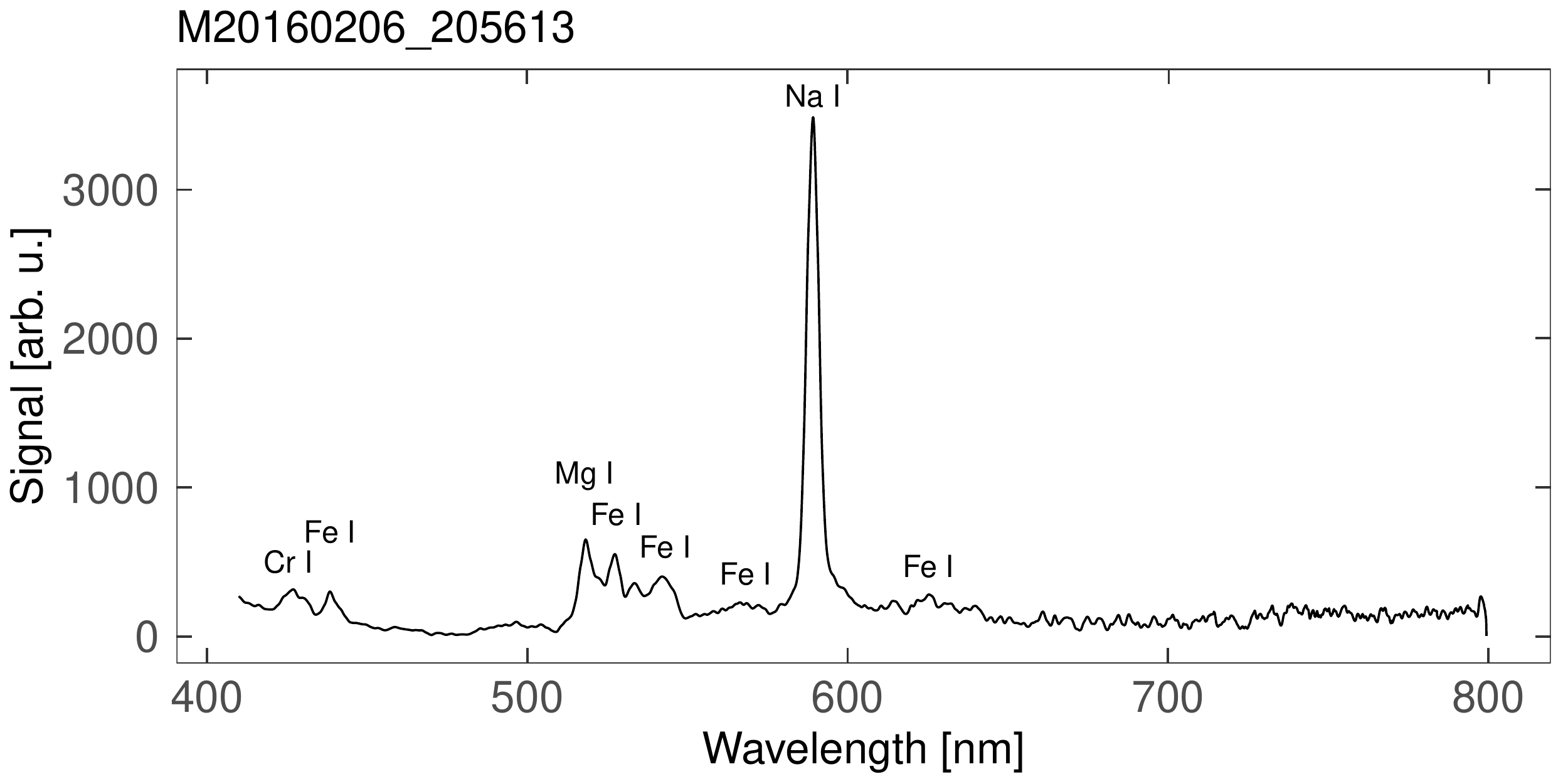}
      \label{fig:1}
    \end{subfigure}%
    ~
    \begin{subfigure}[b]{0.5\textwidth}
      \centering
      \includegraphics[width=\textwidth]{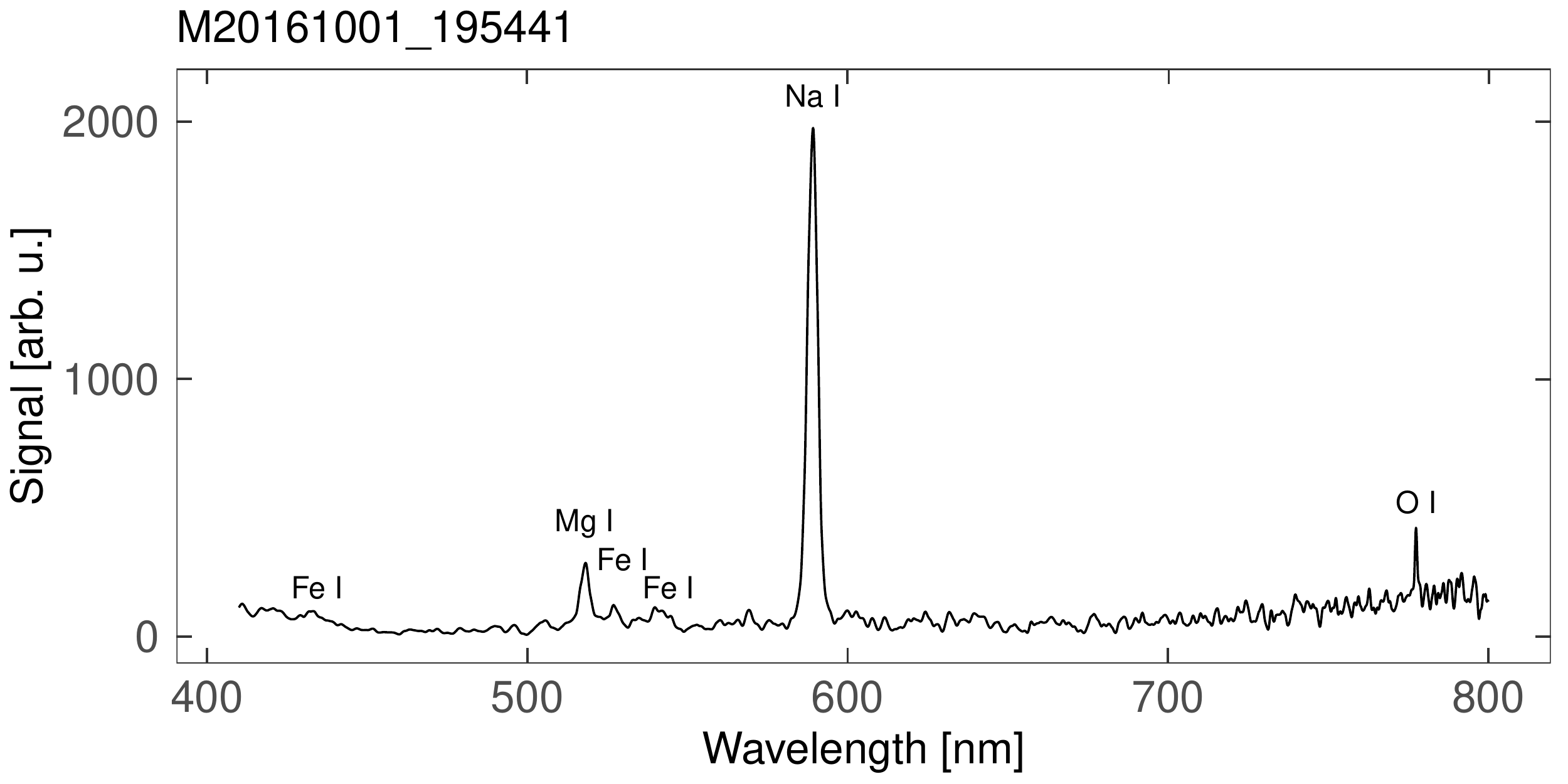}
      \label{fig:2}
    \end{subfigure}%
    \\ 
    \begin{subfigure}[b]{0.5\textwidth}
      \centering
      \includegraphics[width=\textwidth]{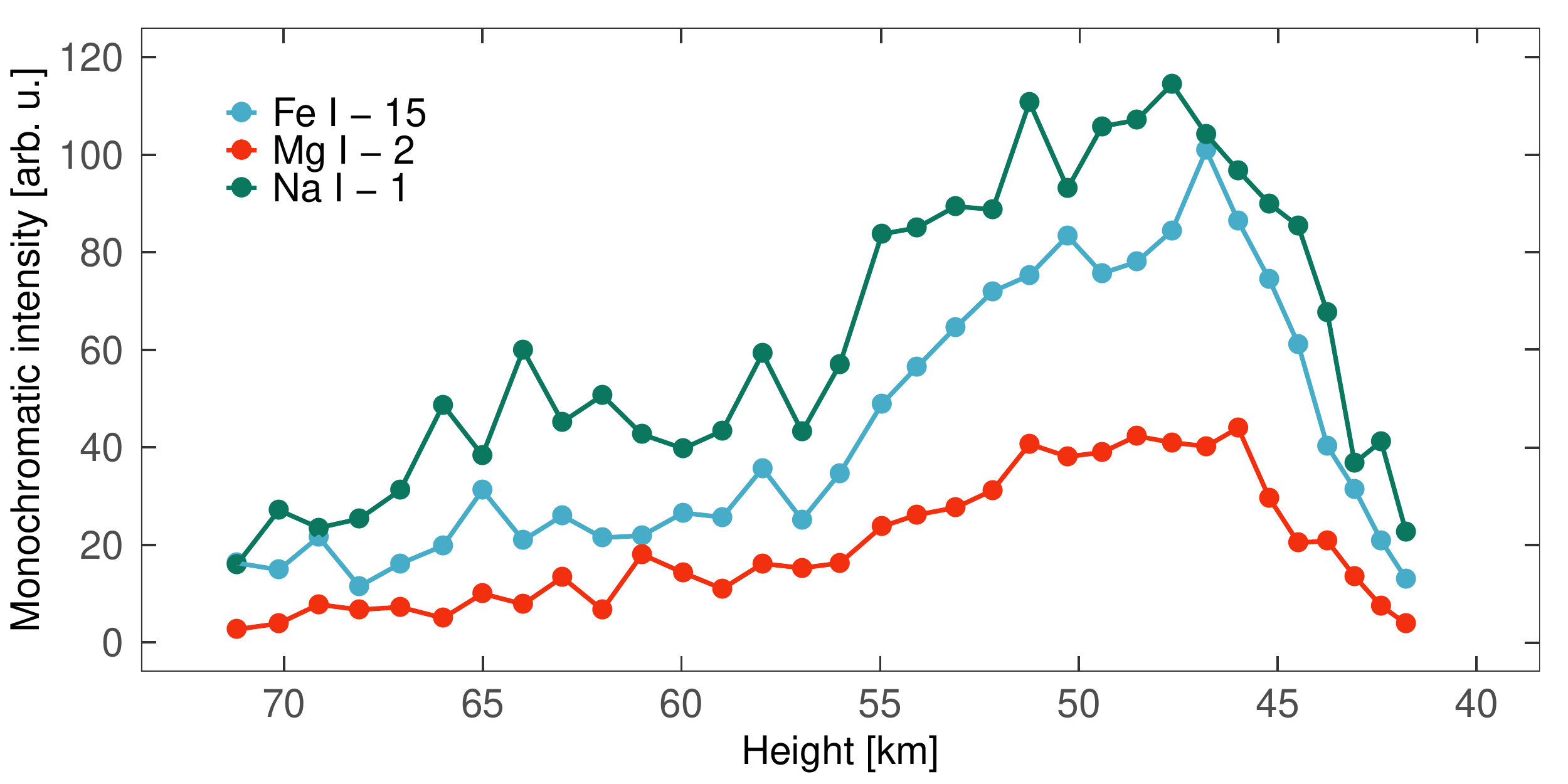}
      \caption{}
      \label{fig:3}
    \end{subfigure}%
    ~
    \begin{subfigure}[b]{0.5\textwidth}
      \centering
      \includegraphics[width=\textwidth]{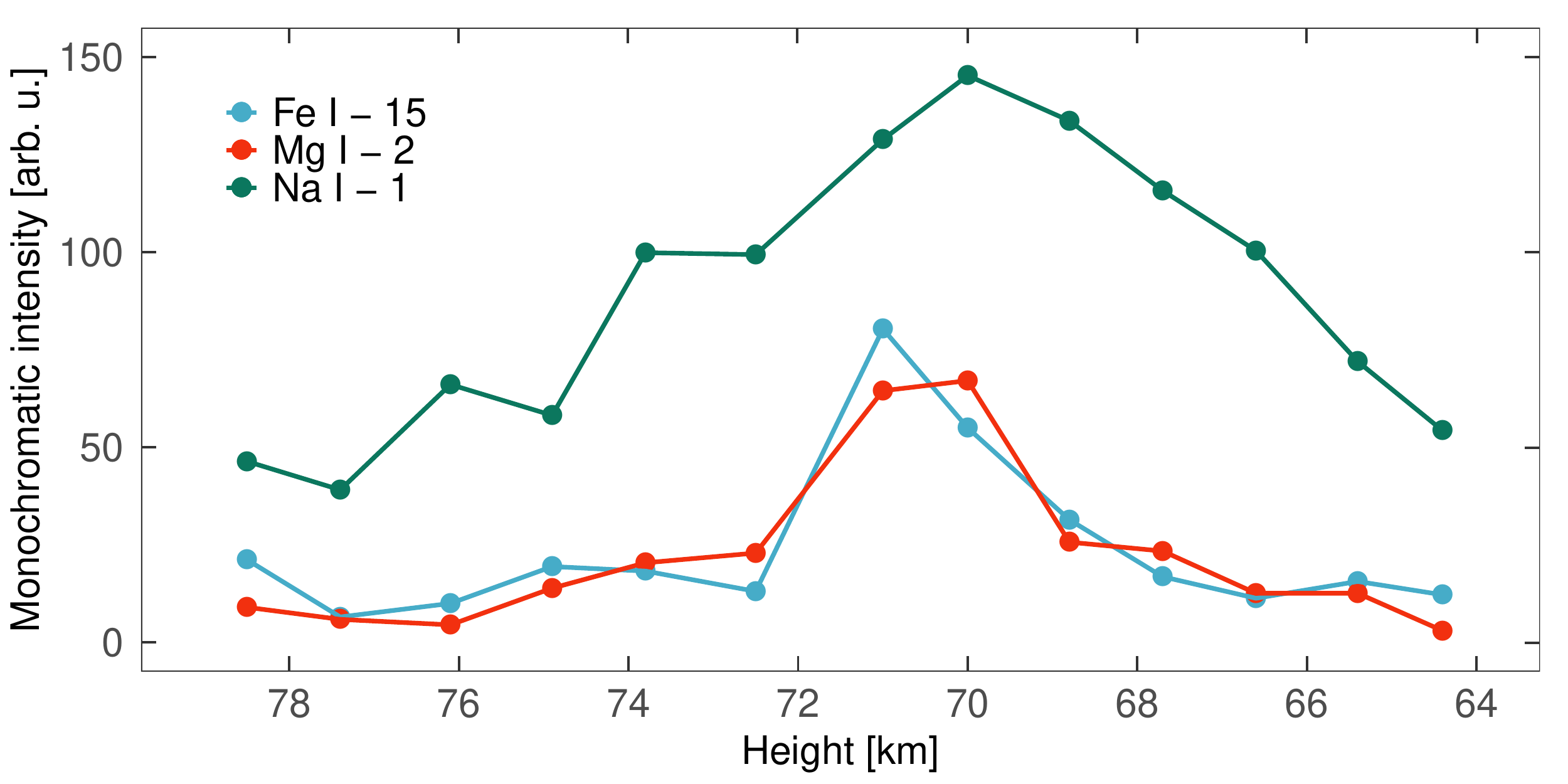}
      \caption{}
      \label{fig:4}
    \end{subfigure}%
    \caption{Summed spectral profile and monochromatic light curves of an asteroidal Na-enhanced meteor M20160206\_205613 (a) and a cometary Na-enhanced meteor M20161001\_195441 (b). Light intensity values below 10 arb. u. are at the noise level of the recording.}
    \label{NaE_LC}
  \end{figure*}

Besides the dominant Na line, the spectra of Na-enhanced meteors consist most notably (Fig. \ref{NaE_LC}) of the Mg I - 2 multiplet near 517.9 nm, the iron multiplets Fe I - 15 (526.0 -- 545.0 nm), Fe I - 41 (438.4 nm) and Fe I - 42 (426.0 -- 436.0 nm) mixed with Cr I - 1 multiplet (peak at 425.5 nm). These lines are among the most prominent also in most normal type meteors. In all cases, Na line begins to radiate early in the higher atmosphere and is also the last to disappear. The orbital and physical properties of the studied Na-enhanced meteors are in Table \ref{strNaE} and \ref{orbNaE}.

\begin{figure*}[!h]
    \centering
    \begin{subfigure}[b]{0.5\textwidth}
      \centering
      \includegraphics[width=\textwidth]{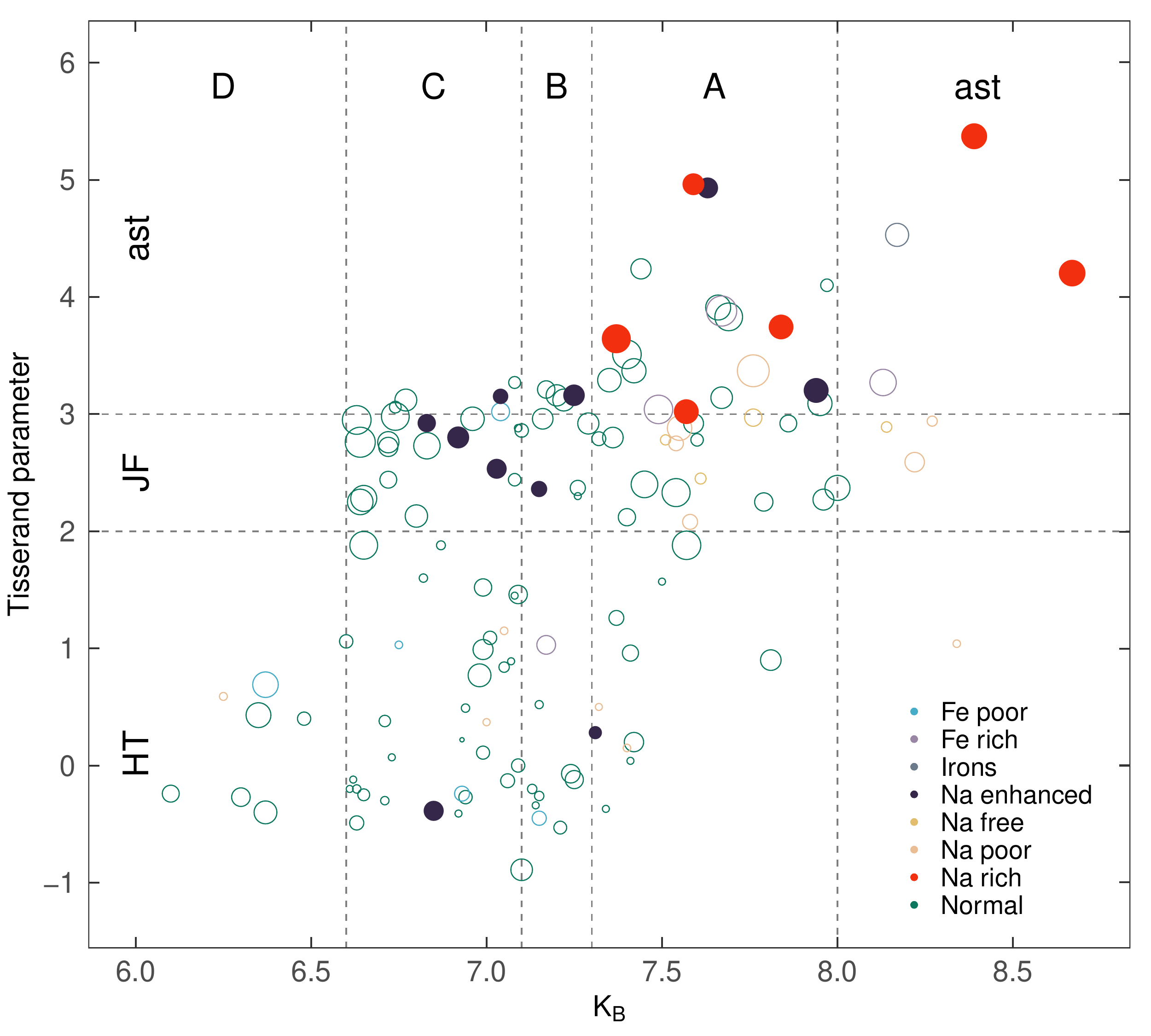}    
      \label{fig:1}
    \end{subfigure}%
    ~
    \begin{subfigure}[b]{0.5\textwidth}
      \centering
      \includegraphics[width=\textwidth]{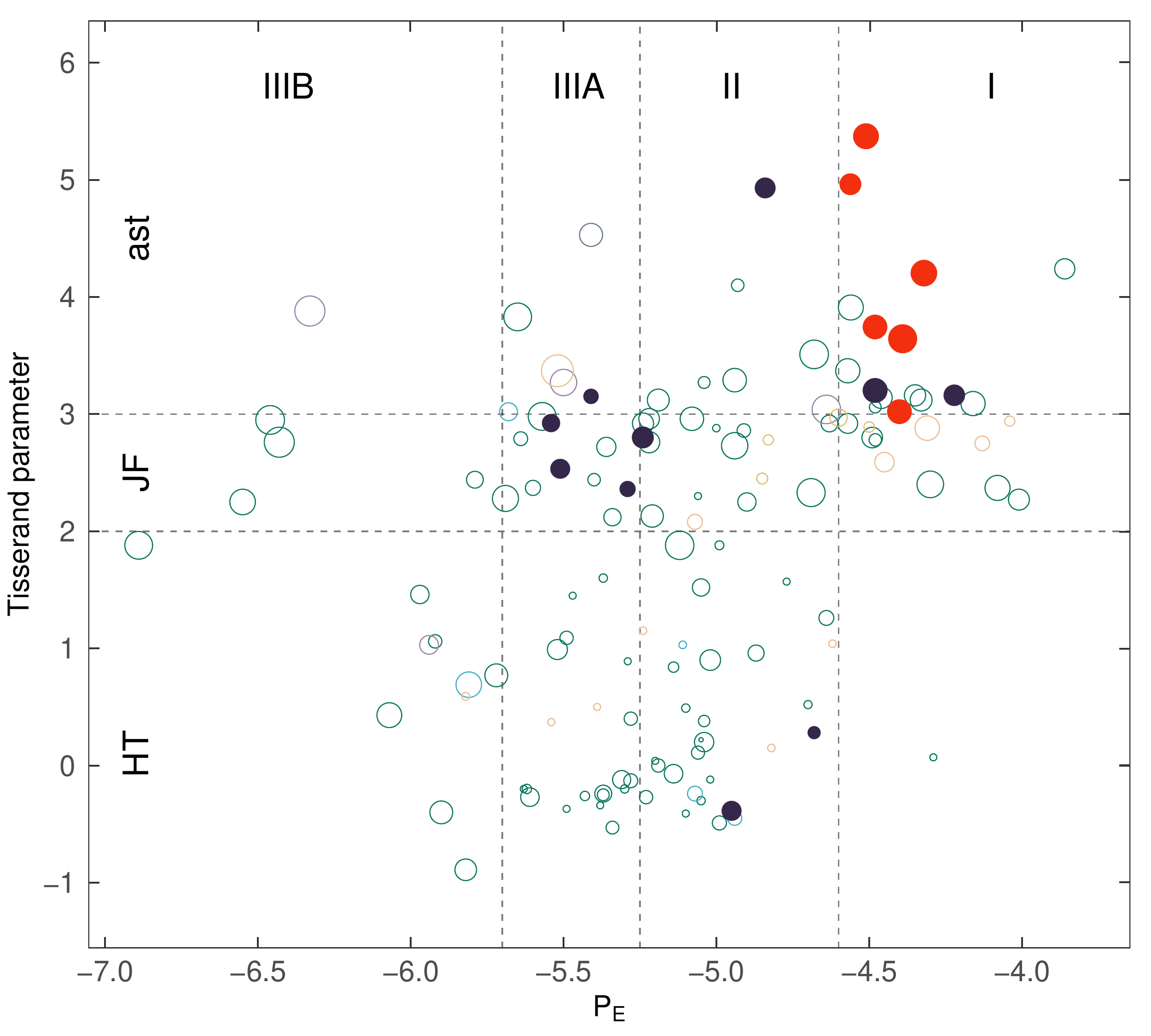}      
      \label{fig:2}
    \end{subfigure}%
    \caption{Material strength classification of meteoroids with Na-enhanced (dark) and Na-rich (red) spectra based on the $K_B$ (left) and $P_E$ (right) parameter as a function of the Tisserand parameter relative to Jupiter. Meteoroids of other spectral types from the survey of \citet{2019A&A...629A..71M} are displayed for comparison. Sizes of Na-enhanced and Na-rich marks reflect relative meteoroid masses in logarithmic scale.}
    \label{KbPe}
  \end{figure*}%
  
Based on their spectra, heliocentric orbits and material strengths, it is clear that Na-enhanced meteors comprise a heterogeneous group with various origin and structure. From spectral point of view, the found heterogeneities relate the variations of Fe intensity, contributions of other faint lines, and the pattern in which these lines emit as a function of height. 

Generally, we can distinguish two types of meteoroids with Na-enhanced spectra. Stronger, asteroidal meteoroids have longer luminous trails and monochoromatic light curves, and larger Fe content (Fig. \ref{NaE_LC}). Based on their material strengths (Table \ref{strNaE}, Fig. \ref{KbPe}), these are on average of type A/I (similar to carbonaceous or ordinary chondrites).

The second group are cometary meteoroids, which are typically of fragile material strength (typically type C/IIIA - standard cometary strength), have shorter luminous trails and lower Fe content. They more often exhibit flares in brightness, during which most of the present emission features are identified.
  
The material heterogeneity correlates with the determined heliocentric orbits (Table \ref{orbNaE}). Stronger meteoroids with Na-enhanced spectra originated on asteroidal orbits with Tisserand parameter $T_J > 3.1$. Cometary Na-enhanced meteors originated mainly from Jupiter-family type or Halley-type orbits. One orbit with $T_J$ = 3.15 among the cometary Na-enhanced meteors belongs to the $\alpha$-Capricornid stream. In previous works, \citet{2005Icar..174...15B} and \citet{2015A&A...580A..67V} identified 11 Na-enhanced meteors, out of which six originated from asteroidal orbits, including two Taurid meteors, four had Jupiter-family type orbits, and one meteor originated in the Ursid stream, derived from Halley-type comet 8P/Tuttle. A degree of Na enhancement among Taurids was also reported by \citet{2017P&SS..143..104M}.

\begin{table*}[!t]
\centering
\small\begin{center}
\caption {Spectral, atmospheric and structural properties of Na-enhanced meteors. Na-enhanced meteoroids in which we report real compositional Na enhancement are marked in italics. Each meteor is designated with characteristic code based on the date and time of observation in UTC (format yyyymmdd\_hhmmss), absolute magnitude (Mag), Na/Mg intensity ratio, Fe/Mg intensity ratio, radiant right ascension $\alpha_g$ in deg, radiant declination $\delta_g$ in deg, initial velocity ($v_i$), beginning height ($H_B$), terminal height ($P_E$), $K_B$ parameter and corresponding material strength class, $P_E$ parameter and corresponding material strength class. Grain density ($\delta_m$) in g\,cm\textsuperscript{-3} is given in the accuracy on the order of the last digit.} 
\vspace{0.1cm}
\resizebox{0.78\textwidth}{!}{\begin{tabular}{lrrrrrrrrrcrcccc}
\hline\hline\\
\multicolumn{1}{c}{Code}& %
\multicolumn{1}{c}{Mag}& %
\multicolumn{1}{c}{Na/Mg} & %
\multicolumn{1}{c}{Fe/Mg} & %
\multicolumn{1}{c}{$\alpha_g$} & %
\multicolumn{1}{c}{$\delta_g$} & %
\multicolumn{1}{c}{$v_i$} & %
\multicolumn{1}{c}{$H_B$} & %
\multicolumn{1}{c}{$H_E$}& %
\multicolumn{1}{c}{$K_B$}& %
\multicolumn{1}{c}{}& %
\multicolumn{1}{c}{$P_E$}& %
\multicolumn{1}{c}{}& %
\multicolumn{1}{c}{$\delta_m$} %
\\
\hline\\
M20140923\_211838 & -1.9 & 7.17  & 1.88 & 344.06 & -2.09  & 21.55 & 93.20  & 72.69 & 7.15 & B  & -5.29 & IIIA & 2.3 \\
				  & 1.1  & 1.19  & 0.42 & 0.01   & 0.04   & 0.10  & 0.09   & 0.09  & 0.02 &    & 0.20  &  &\vspace{0.1cm}\\
				  
\textit{M20150724\_213217} & -2.4 & 10.07 & 4.46 & 305.41 & -10.52 & 26.52 & 96.72  & 80.43 & 7.04 & C1 & -5.41 & IIIA & 2.1 \\
				  & 0.5  & 2.48  & 1.22 & 0.47   & 0.20   & 0.11  & 0.11   & 0.08  & 0.03 &    & 0.09  &  &\vspace{0.1cm}\\
				  
\textit{M20150811\_012547} & -3.0 & 6.44  & 3.36 & 43.22  & 56.75  & 59.55 & 103.85 & 80.97 & 7.31 & A  & -4.68 & II   & 3.0 \\
 				  & 1.0  & 4.65  & 2.72 & 0.52   & 0.12   & 0.77  & 0.34   & 0.13  & 0.01 &    & 0.17  &  &\vspace{0.1cm}\\
 				  
\textit{M20150908\_210450} & -3.6 & 7.08  & 0.99 & 258.79 & 56.56  & 22.82 & 94.51  & 71.79 & 7.03 & C1 & -5.51 & IIIA & 2.0 \\
				  & 0.5  & 0.52  & 0.13 & 0.81   & 0.37   & 0.15  & 0.13   & 0.17  & 0.02 &    & 0.09  &  &\vspace{0.1cm}\\
				  
M20160907\_205714 & -1.9 & 10.04 & 4.29 & 273.66 & 30.55  & 16.22 & 92.51  & 70.09 & 6.83 & C1 & -5.54 & IIIA & 2.1 \\
				  & 1.1  & 3.54  & 2.11 & 0.12   & 0.07   & 0.22  & 0.05   & 0.04  & 0.02 &    & 0.11  &  &\vspace{0.1cm}\\

M20161001\_195441 & -3.3 & 7.33  & 1.47 & 326.84 & 16.02  & 17.53 & 92.49  & 60.42 & 6.92 & C1 & -5.24 & II   & 1.8 \\
				  & 1.0  & 0.62  & 0.18 & 0.07   & 0.03   & 0.19  & 0.03   & 0.01  & 0.02 &    & 0.18  &  &\vspace{0.1cm}\\

\textit{M20161002\_013415} & -6.7 & 3.18  & 1.13 & 82.64 & 67.52  & 59.58  & 105.01 & 79.15  & 7.32 & A & -5.56 & IIIA & 3.7 \\
			      & 1.5  & 0.08  & 0.13 & 0.97  & 0.41   & 0.33   & 0.20   & 0.16   & 0.03 &    & 0.24 & \vspace{0.1cm} \\
			      				  
\textit{M20170801\_204124} & -6.7 & 4.21  & 1.09 & 37.89  & 56.40  & 60.02 & 112.04 & 77.64 & 6.85 & C2 & -4.95 & II   & 2.2 \\
				  & 1.3  & 0.42  & 0.17 & 0.40   & 0.03   & 0.65  & 0.49   & 0.03  & 0.02 &    & 0.23  &  &\vspace{0.2cm}\\
				  
M20160811\_212632 & -4.2 & 7.91  & 2.33 & 297.90 & -17.72 & 17.55 & 81.67  & 48.63 & 7.94 & A  & -4.48 & I    &  \\
 				  & 0.7  & 0.83  & 0.31 & 0.02   & 0.12   & 0.30  & 0.05   & 0.11  & 0.02 &    & 0.32  &  &\vspace{0.1cm}\\
 				  
M20161002\_002244 & -1.8 & 9.27  & 2.59 & 5.98   & -16.34 & 14.66  & 82.01  & 57.27 & 7.63 & A  & -4.84 & II   &  \\
				  & 1.0  & 3.52  & 2.66 & 0.01   & 0.13   & 0.16   & 0.09   & 0.03  & 0.02 &    & 0.18 &  &\vspace{0.1cm}\\
				  
M20170402\_020952 & -2.5 & 7.51  & 2.15 & 10.74  & 72.36  & 18.30 & 90.48  & 45.77 & 7.25 & B  & -4.22 & I    &  \\
 				  & 0.7  & 0.79  & 0.29 & 0.09   & 0.06   & 0.12  & 0.09   & 0.04  & 0.02 &    & 0.10  &  &\vspace{0.1cm}\\
\hline
\end{tabular}} 
\label{strNaE}
\end{center}
\end{table*}

\begin{table*}[!h]
\centering
\small\begin{center}
\caption {Orbital properties of Na-enhanced meteors. Na-enhanced meteoroids in which we report real compositional Na enhancement are marked in italics. Each meteor is designated with a code as in Table \ref{strNaE}, Type differentiating between cometary (C) and asteroidal (A), semi-major axis $a$ in au, eccentricity $e$, perihelion distance $q$ in au, aphelion distance $Q$ in au, inclination $i$ in deg, argument of perihelion $\omega$ in deg, longitude of the ascending node $\Omega$ in deg, Tisserand parameter with respect to Jupiter ($T_J$) and corresponding orbit type, and associated meteor shower in the IAU MDC code.} 
\vspace{0.1cm}
\resizebox{0.78\textwidth}{!}{\begin{tabular}{lcrrrrrrrccccc}
\hline\hline\\
\multicolumn{1}{c}{Code}& %
\multicolumn{1}{c}{Type}& %
\multicolumn{1}{c}{$a$} & %
\multicolumn{1}{c}{$e$} & %
\multicolumn{1}{c}{$q$} & %
\multicolumn{1}{c}{$Q$} & %
\multicolumn{1}{c}{$i$} & %
\multicolumn{1}{c}{$\omega$}& %
\multicolumn{1}{c}{$\Omega$}& %
\multicolumn{1}{c}{$T_J$}& %
\multicolumn{1}{c}{Type}& %
\multicolumn{1}{c}{Shower} %
\\
\hline\\
M20140923\_211838 & C & 3.91   & 0.81 & 0.762 & 7.06   & 2.30   & 242.61 & 180.57  & 2.36  & JF  & spo   \\
				  &   & 0.14   & 0.02 & 0.005 & 0.33   & 0.20   & 0.12   &         & 0.08  &     &  \vspace{0.1cm}\\
				  
\textit{M20150724\_213217} & C & 2.24   & 0.77 & 0.516 & 3.97   & 7.44   & 277.60 & 121.45  & 3.15  & AST & CAP   \\
			      &   & 0.07   & 0.02 & 0.004 & 0.17   & 0.26   & 0.89   &         & 0.09  &     &  \vspace{0.1cm}\\
			      
\textit{M20150811\_012547} & C & 7.03   & 0.86 & 0.960 & 13.10  & 113.35 & 152.31 & 137.88  & 0.28  & HT  & PER   \\
			      &   & 5.47   & 0.06 & 0.004 & 10.59  & 0.54   & 1.45   &         & 0.70  &     &  \vspace{0.1cm}\\
			      
\textit{M20150908\_210450} & C & 3.38   & 0.70 & 1.007 & 5.75   & 30.64  & 176.99 & 165.68  & 2.53  & JF  & NDR   \\
				  &   & 0.21   & 0.03 & 0.008 & 0.46   & 0.27   & 0.64   &         & 0.10  &     & \vspace{0.1cm} \\
				  
M20160907\_205714 & C & 2.58   & 0.61 & 1.000 & 4.16   & 15.06  & 191.45 & 165.42  & 3.09  & AST  & spo   \\
				  &   & 0.08   & 0.03 & 0.002 & 0.13   & 0.48   & 0.24   &         & 0.05  &     & \vspace{0.1cm} \\
				  
M20161001\_195441 & C & 3.01   & 0.70 & 0.909 & 5.12   & 9.78   & 218.95 & 188.84  & 2.80  & JF  & spo   \\
			      &   & 0.15   & 0.02 & 0.003 & 0.32   & 0.20   & 0.06   &         & 0.11  &     & \vspace{0.1cm} \\
			      
\textit{M20161002\_013415} & C & -      & 1.12 & 0.951 & 105.24 & -      & 205.02 & 189.06  & -     & HT  & spo   \\
			      &   & -      & 0.15 & 0.003 & 0.96   & -      & 0.54   &         & -     &     & \vspace{0.1cm} \\
			      
\textit{M20170801\_204124} & C & 111.18 & 0.99 & 0.922 & 221.44 & 111.49 & 144.71 & 129.538 & -0.39 & HT  & PER   \\
  				  &   & -      & 0.05 & 0.006 & -      & 0.44   & 1.63   &         & 0.63  &     & \vspace{0.2cm} \\
  				  
M20160811\_212632 & A & 2.31   & 0.63 & 0.854 & 3.77   & 1.13   & 233.73 & 139.35  & 3.29  & AST & spo   \\
				  &   & 0.12   & 0.03 & 0.006 & 0.27   & 0.27   & 0.22   &         & 0.12  &     & \vspace{0.1cm} \\
				  
M20161002\_002244 & A & 1.30   & 0.37 & 0.823 & 1.77   & 5.29   & 70.28  & 9.01    & 4.93  & AST & spo   \\
				  &   & 0.08   & 0.02 & 0.004 & 0.14   & 0.09   & 0.48   &         & 0.21  &     & \vspace{0.1cm} \\
				  
M20170402\_020952 & A & 2.44   & 0.61 & 0.965 & 3.92   & 19.43  & 155.40 & 12.29   & 3.16  & AST & spo   \\
				  &   & 0.21   & 0.05 & 0.003 & 0.47   & 0.16   & 0.14   &         & 0.18  &     &  \\
\hline
\end{tabular}} 
\label{orbNaE}
\end{center}
\end{table*}

Finally, we estimated grain (mineralogical) densities of the cometary meteoroids with Na-enhanced spectra based on the heat-conductivity method \citep{1958BAICz...9..154C, 2009A&A...495..353B}. The obtained values ranged from 1.8 to 3.7 g\,cm\textsuperscript{-3} with mean value of 2.3 $\pm$ 0.6 g\,cm\textsuperscript{-3}. The low grain densities are consistent with the fragile cometary material $strengths$ of this group. The obtained values are comparable to other Perseid or even Draconid meteoroids originating from comet 109P/Swift-Tuttle and 21P/Giacobini-Zinner respectively \citep{2009A&A...495..353B}. Meteoroids with low grain density may suggest presence of carbonaceous material and could also be connected to higher porosity. One low density Na-enhanced meteor studied by \citet{2008EM&P..102..485B} was explained by structure consisting of fewer and larger grains. 

The grain densities of asteroidal Na-enhanced and Na-rich meteors yielded scattered results, which were regarded as unreliable. The method is highly dependent on the meteor beginning height and initial speed, and is in given sensitivity of the camera probably not suitable for very slow meteors.

\subsection{Na-rich meteors} \label{NaRich}

Na-rich meteors are defined as having Na/Mg and Na/Fe ratios obviously higher than expected for chondritic meteoroids \citep{2005Icar..174...15B}. The detection of the main Mg and Fe emission lines was difficult in two of the six analyzed meteors. In other cases, they were present near the noise level of the spectrum. It is probable that the detection of these lines is complicated by the low temperature of the radiating gas and insufficient S/N ratio.

\begin{figure}[t]
\centerline{\includegraphics[width=9cm,angle=0]{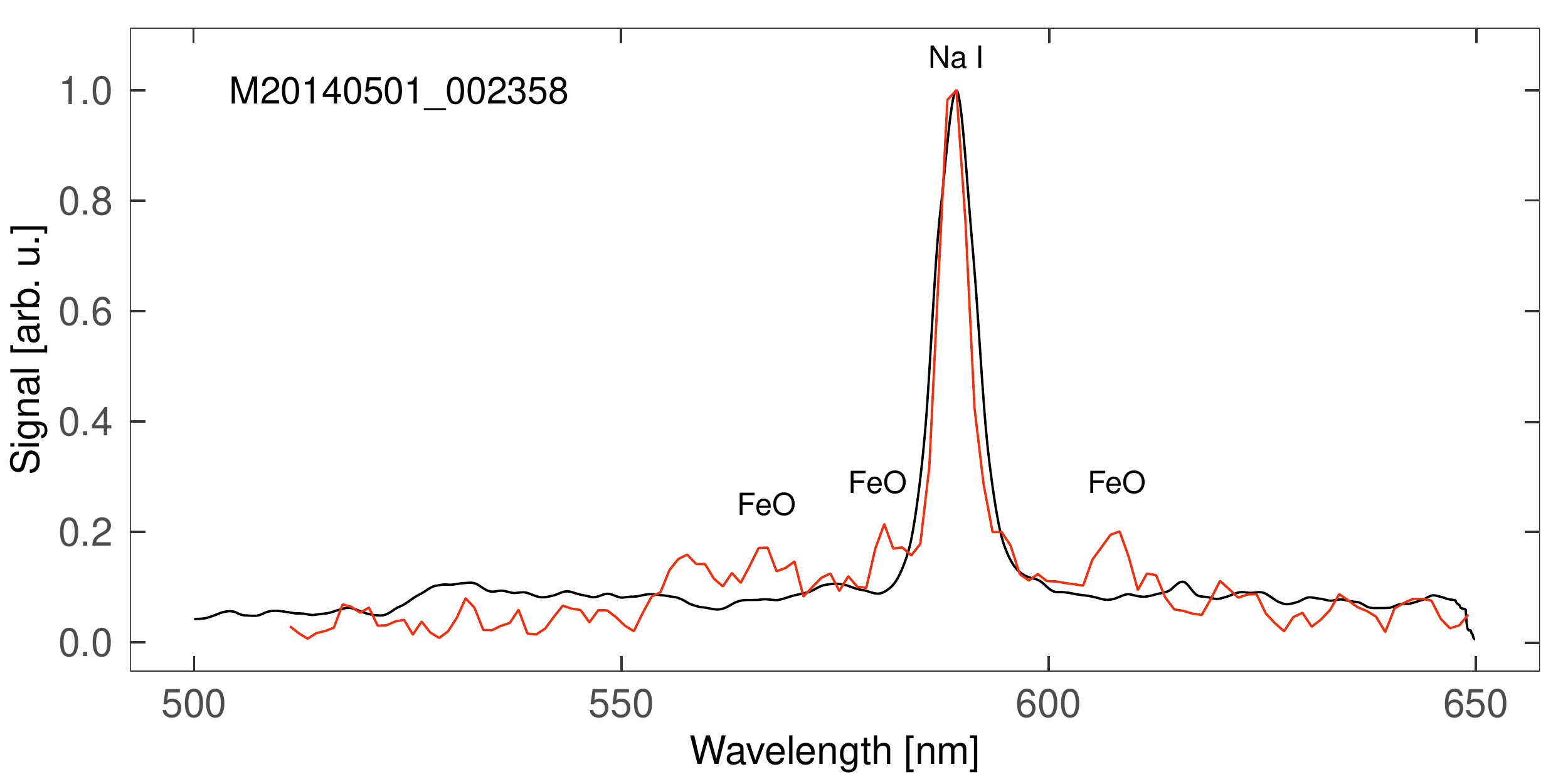}}
\centerline{\includegraphics[width=9cm,angle=0]{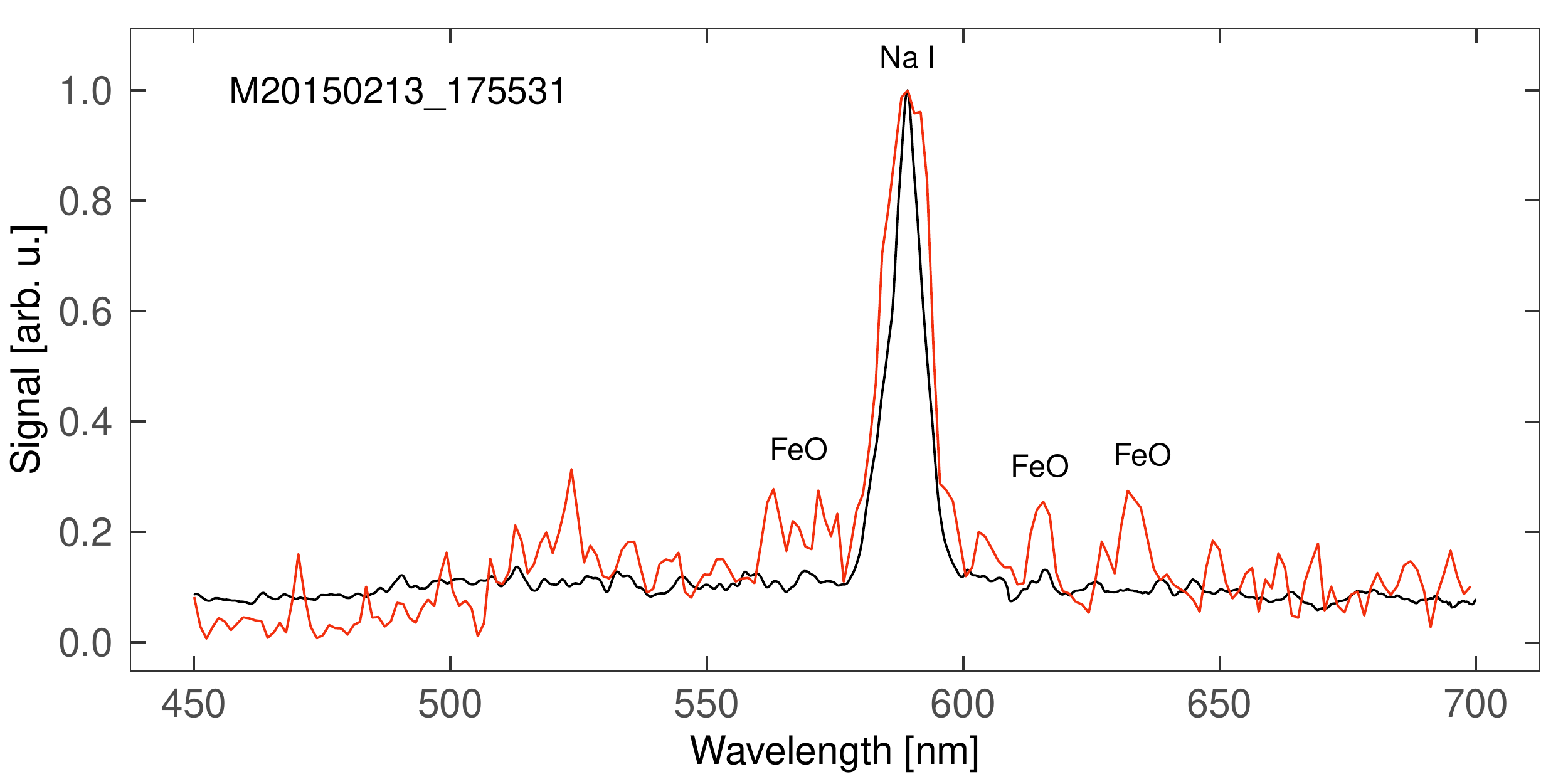}} \caption[f1]{Summed spectral profiles (black) of two Na-rich meteors along with spectrum of a short subtle flare (red) captured in one frame in each recording. Signal in arbitrary units is for comparison normalized to unity at the peak intensity of the 589.2 nm Na line.} 
\label{NaR_flare}
\end{figure}

Some Na-rich meteors exhibited short subtle flares during their ablation. In these events (red curves in Fig. \ref{NaR_flare}), faint spectral features were observed near the dominant Na line at 560 - 581 nm and 608 - 630 nm. These features were most likely the result of the FeO band emission.

It is assumed that FeO can be ablated directly in molecular form at high altitudes \citep{2016Icar..278..248B}. FeO and other oxides are also expected to form significantly after cooling at high pressures corresponding to low altitudes. Characteristic temperature of this process is expected in the 2000 - 2500 K range, which could correspond to the lower achieved temperatures of slow Na-rich meteors. Based on our data, FeO emission is the second most important contribution in the studied Na-rich spectra.

\begin{table*}[t]
\centering
\small\begin{center}
\caption {Spectral, atmospheric and structural properties of Na-rich meteors. Given parameters are the same as in Table \ref{strNaE}.} 
\vspace{0.1cm}
\resizebox{0.7\textwidth}{!}{\begin{tabular}{lrrrrrrrrrcrlclc}
\hline\hline\\
\multicolumn{1}{c}{Code}& %
\multicolumn{1}{c}{Mag}& %
\multicolumn{1}{c}{Na/Mg} & %
\multicolumn{1}{c}{Fe/Mg} & %
\multicolumn{1}{c}{$\alpha_g$} & %
\multicolumn{1}{c}{$\delta_g$} & %
\multicolumn{1}{c}{$v_i$} & %
\multicolumn{1}{c}{$H_B$} & %
\multicolumn{1}{c}{$H_E$}& %
\multicolumn{1}{c}{$K_B$}& %
\multicolumn{1}{c}{}& %
\multicolumn{1}{c}{$P_E$}& %
\multicolumn{1}{c}{} %
\\
\hline\\
M20140501\_002358 & -3.4 & 63.66 & 7.12 & 193.58 & -43.89 & 13.75 & 73.86 & 54.06 & 8.39 & ast & -4.51 & I  \\
			      & 0.7  & 12.08 & 2.42 & 0.38   & 1.04   & 0.29  & 1.27  & 0.65  & 0.03 &     & 0.14  & \vspace{0.1cm} \\

M20150213\_175531 & -1.7 & 39.98 & 3.67 & 115.94 & -3.63  & 13.19 & 79.58 & 47.41 & 7.59 & A   & -4.56 & I  \\
				  & 0.6  & 11.07 & 1.26 & 0.17   & 0.26   & 0.19  & 0.12  & 0.03  & 0.02 &     &  0.12 & \vspace{0.1cm} \\

M20150601\_220409 & -3.8 & 12.42 & 1.91 & 84.71  & 27.18  & 16.44 & 76.51 & 61.21 & 8.67 & ast & -4.32 & I  \\
				  & 0.8  & 1.23  & 0.27 & 0.19   & 0.25   & 0.21  & 0.07  & 0.03  & 0.02 &     &  0.15 & \vspace{0.1cm} \\

M20161128\_222137 & -5.5 & 29.43 & 3.76 & 336.87 & -3.48  & 12.70 & 83.55 & 36.34 & 7.37 & A   & -4.40 & I  \\
 				  & 1.0  & 11.84 & 1.36 & 0.20   & 0.18   & 0.13  & 0.06  & 0.31  & 0.02 &     &  0.18 & \vspace{0.1cm} \\

M20171029\_234653 & -3.1 & 34.33 & 4.81 & 325.90 & -15.82 & 13.70 & 81.98 & 41.86 & 7.57 & A   & -4.39 & I  \\
 				  & 0.9  & 12.41 & 2.04 & 0.13   & 0.13   & 0.11  & 0.08  & 0.05  & 0.02 &     &  0.16 & \vspace{0.1cm} \\

M20180924\_221936 & -4.2 & 14.09 & 2.47 & 343.30 & -4.25  & 16.94 & 80.57 & 44.55 & 7.84 & A   & -4.48 & I  \\
 				  & 0.2  &  1.95 & 0.45 & 0.07   & 0.07   & 0.12  & 0.04  & 0.06  & 0.02 &     &  0.06 & \vspace{0.1cm} \\ 				  
\hline \vspace{0.4cm}
\end{tabular}} 
\label{strNaR}
\end{center}
\end{table*}

\begin{table*}[!h]
\centering
\small\begin{center}
\caption {Orbital properties of Na-rich meteors. Given parameters are the same as in Table \ref{orbNaE}.} 
\vspace{0.1cm}
\resizebox{0.6\textwidth}{!}{\begin{tabular}{lrrrrrrrrrc}
\hline\hline\\
\multicolumn{1}{c}{Code}& %
\multicolumn{1}{c}{$a$} & %
\multicolumn{1}{c}{$e$} & %
\multicolumn{1}{c}{$q$} & %
\multicolumn{1}{c}{$Q$} & %
\multicolumn{1}{c}{$i$} & %
\multicolumn{1}{c}{$\omega$}& %
\multicolumn{1}{c}{$\Omega$}& %
\multicolumn{1}{c}{$T_J$}& %
\multicolumn{1}{c}{Type}& %
\multicolumn{1}{c}{Shower} %
\\
\hline\\
M20140501\_002358 & 1.17 & 0.24 & 0.885 & 1.45  & 8.85 & 67.96  & 220.38 & 5.36 & AST & spo  \\
 				  & 0.10 & 0.08 & 0.004 & 0.23  & 0.10 & 0.32   &        & 0.25 &  &  \vspace{0.1cm} \\
                    
M20150213\_175531 & 1.30 & 0.28 & 0.930 & 1.67  & 4.90 & 42.33  & 144.48 & 4.96 & AST & spo  \\
 				  & 0.08 & 0.05 & 0.005 & 0.17  & 0.11 & 0.19   &        & 0.20 &  &  \vspace{0.1cm}\\
                    
M20150601\_220409 & 1.61 & 0.49 & 0.816 & 2.40  & 1.45 & 114.15 & 70.82  & 4.20 & AST & spo  \\
 				  & 0.10 & 0.04 & 0.004 & 0.22  & 0.15 & 0.26   &        & 0.16 &  &  \vspace{0.1cm}\\
                    
M20161128\_222137 & 2.02 & 0.51 & 0.986 & 3.06  & 1.04 & 181.43 & 247.02 & 3.65 & AST & spo  \\
			      & 0.08 & 0.03 & 0.003 & 0.18  & 0.08 & 0.21   &        & 0.09 &  &  \vspace{0.1cm}\\
                    
M20171029\_234653 & 2.74 & 0.64 & 0.986 & 4.49  & 0.46 & 11.19  & 36.31  & 3.01 & AST & spo  \\
 				  & 0.11 & 0.05 & 0.002 & 0.32  & 0.11 & 0.18   &        & 0.05 &  &  \vspace{0.1cm}\\
                    
M20180924\_221936 & 1.90 & 0.56 & 0.827 & 2.97  & 1.00 & 239.05 & 181.57 & 3.74 & AST & spo  \\
				  & 0.09 & 0.04 & 0.002 & 0.22  & 0.08 & 0.07   &        & 0.14 &  &  \vspace{0.1cm}\\
\hline 
\end{tabular}} 
\label{orbNaR}
\end{center}
\end{table*}

\begin{figure*}[h]
\centerline{\includegraphics[width=0.7\textwidth,angle=0]{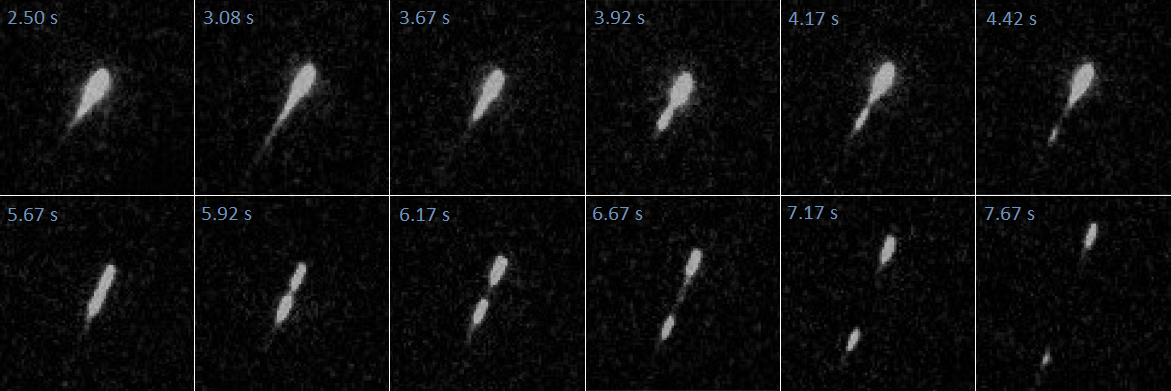}} \caption[f1]{Mass-loss and fragmentation phases of a Na-rich meteor M20150601\_220402. Two events of non-explosive disruption into equal fragments were observed, showing identical spectral features.} 
\label{NaFrag}
\end{figure*}

For further insight into the spectral properties of Na-rich meteors, we supplemented our sample with two higher-resolution spectra captured by the AMOS-Spec-HR system in La Palma, Canary Islands. As a result of the higher sensitivity in the near-UV region, we were able to identify the Ca II lines (H and K doublet) at 393.4 and 396.9 nm. The Ca II doublet is often observed as a primary feature in fast meteors, however, since it is of relatively low excitation (3.15 eV), it can be present in the low-temperature main spectrum component \citep{1994P&SS...42..145B}. The emission of the molecular FeO was not apparent in the higher-resolution Na-rich spectrum.

All Na-rich meteors show high material strength of asteroidal debris (Table \ref{strNaR}), classified based on the $K_B$ and $P_E$ parameters as type ast-A/I. This classification suggests that both carbonaceous and ordinary chondrites could contribute to this group.

Monochromatic light curves of Na-rich meteors are smooth, with only very short and subtle flares in few cases. Sodium starts to burn early and continues steadily throughout the long luminous trajectories. The slow and gradual release of sodium corresponds with the overall light curve. 

Specific fragmentation patterns were detected among the Na-rich meteors, which typically progress in two stages (Fig. \ref{NaFrag}). Upon gradual heating in the upper atmosphere, meteoroid starts to form an elongated luminous wake. The wake could reflect gradual release of dust through quasi-continuous fragmentation \citep[e.g.][]{2002A&A...384..317B} or can be produced by prolonged radiation by rarefied gas at higher altitudes. In time, the meteor wake decreases and meteoroid undergoes gradual fragmentation into large fragments of comparable size, which continue in gradual decay up until the lower atmosphere.

For some meteors, we were able to simultaneously obtain spectra of the two large fragments shortly after their disruption. The spectra showed the same spectral feature -- strong isolated Na line. This suggests no significant heterogeneities are present in the meteoroid composition, though the observed spectra are limited by the generally low S/N ratio.

The very long meteor trails and fragmentation patterns reflect the stronger chondritic material of these bodies. The two events of large-scale disruption into similar fragments in meteoroid M20161128\_222137 occurred at heights of approximately 49.5 km and 41.5 km respectively. The dynamic pressure \citep{Fadeenko} causing the large-scale disruption events was determined at 0.10 $\pm$ 0.03 and 0.21 $\pm$ 0.04 MPa respectively. Such low values suggest that the pre-entry meteoroid was possibly fractured. Data from instrumentally observed meteorite falls have shown that interplanetary meteoroids exhibit very low atmospheric fragmentation strengths in comparison to measured tensile strengths of meteorites \citep{2011M&PS...46.1525P, 2014M&PS...49..328G}.

The determined heliocentric orbits of all six Na-rich meteors also indicate asteroidal origin (Table \ref{orbNaR}). With semi-major axes between 1.17 and 2.70 au and aphelion distance 1.45 $< Q <$ 4.50 au, these meteors had typical Apollo-type orbits. Previous surveys of smaller mm-sized meteoroids \citep{2005Icar..174...15B, 2019A&A...621A..68V} provided overall three Na-rich meteors with heliocentric orbits, all classified as Jupiter-family type. 

Furthermore, we have applied the Soutworth-Hawkins \citep{1963SCoA....7..261S} and Drummond \citep{1981Icar...45..545D} orbital similarity criteria to look for possible associated parent objects of individual meteoroids. For most of the Na-rich and asteroidal Na-enhanced meteors, we have found several Apollo asteroids satisfying criteria for possible association ($D_{SH} <$ 0.08 and $D_{D} <$ 0.05). The found similarities may however be only accidental, caused by the dense NEO space in this region. To confirm association with any of the potential parent asteroids would require further dynamical and statistical analysis \citep[e.g.][]{2019A&A...622A..84G, 2004EM&P...95..697P}, which is beyond the scope of this work. 

The found Apollo asteroids on orbits close to studied meteoroids are typically of 23 to 29 absolute magnitude, which assuming different albedos corresponds to sizes of approximately 4 - 150 m in diameter. Bodies of such size are probably not capable of producing sufficiently large dust clouds for the formation of meteoroid streams. We rather assume the production of individual mm- to dm-sized fragments created by impact or past large scale disruption. Furthermore, the recently discovered ejection of cm- to dm-sized particles from the surface of Bennu \citep{2019Natur.568...55L, 2019DDA....5010005M, KovacovaBennu} raises the possibility for another mechanism causing the release of mid-sized meteoroids from asteroids.

\section{Speed dependency and spectra of slow meteors} \label{SpeedDep}

\begin{figure*}[t]
\begin{center}
\includegraphics[width=.45\textwidth]{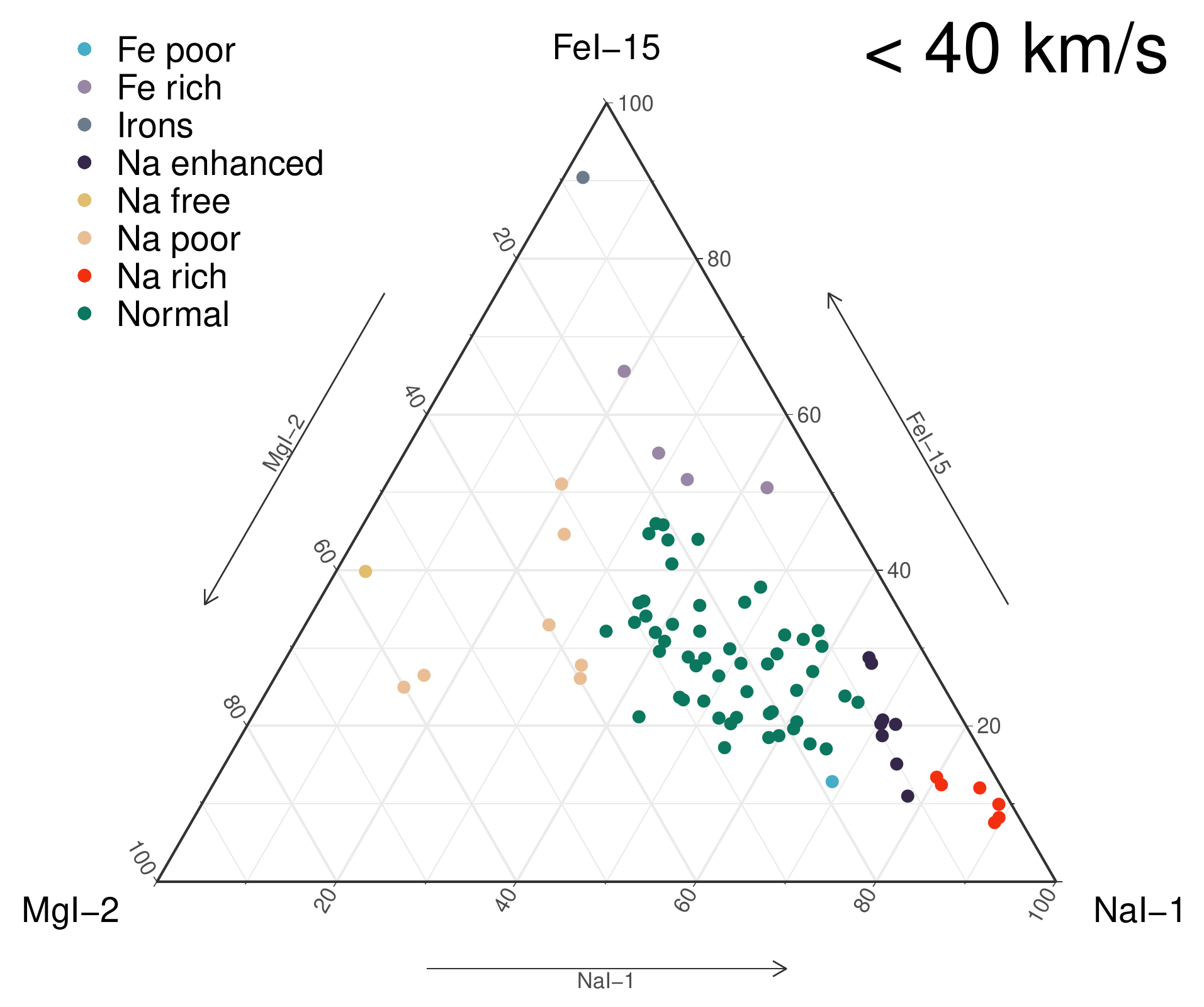}
\includegraphics[width=.45\textwidth]{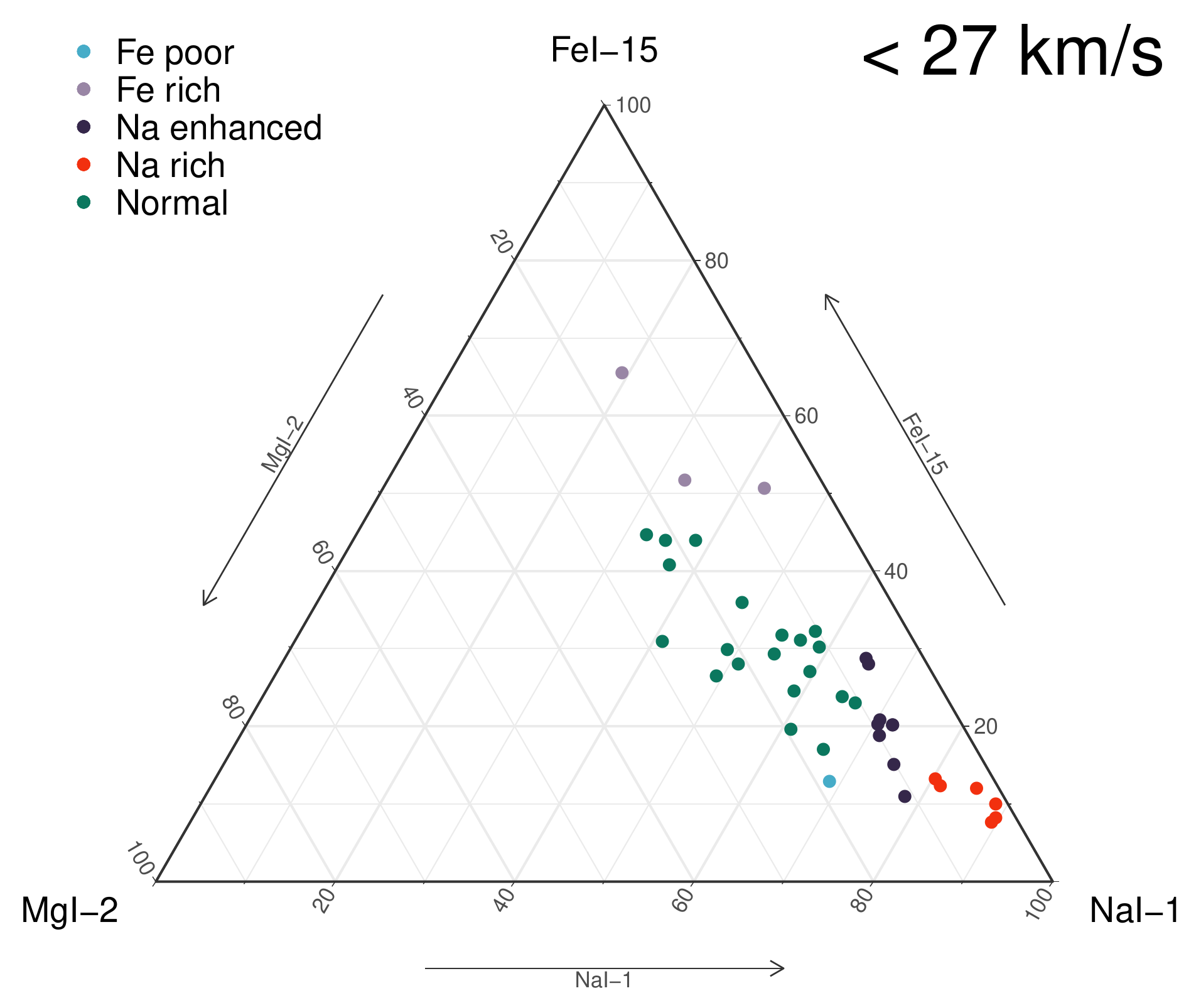}
\includegraphics[width=.45\textwidth]{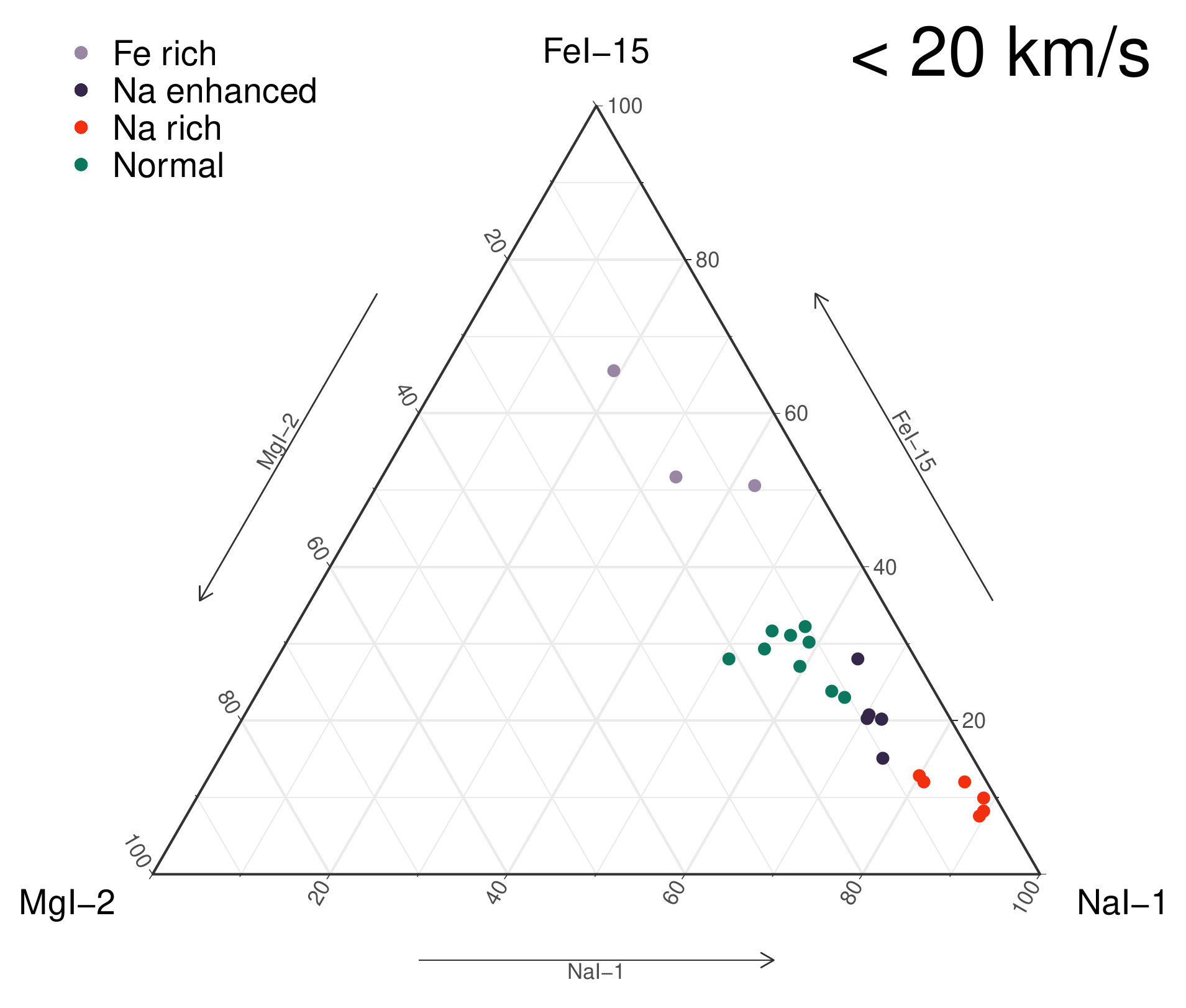}
\includegraphics[width=.45\textwidth]{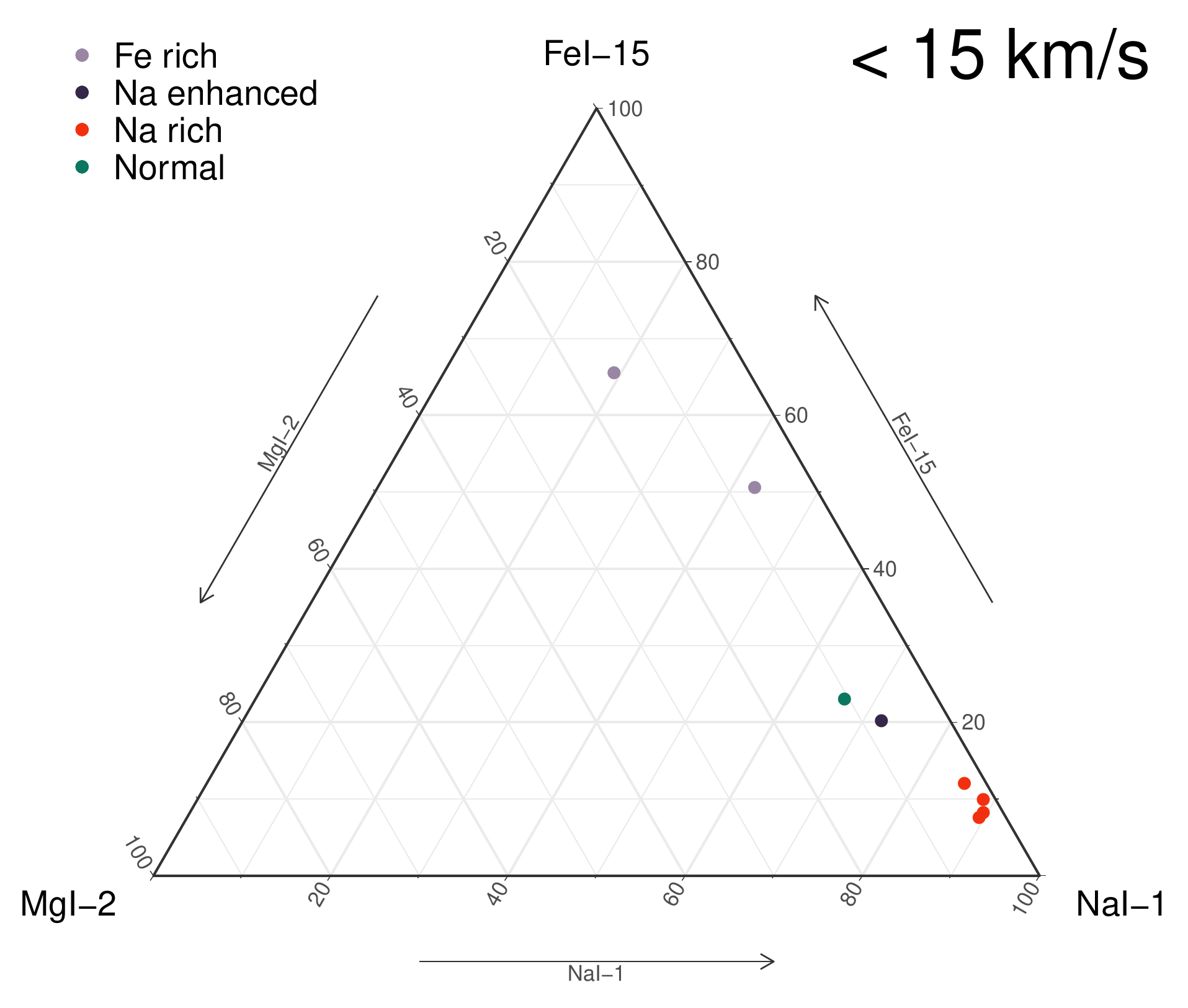}
\end{center}
\caption{Spectral classification based on relative intensity ratios of Mg I - 2, Na I - 1 and Fe I - 15 for meteors with different upper threshold values of entry meteor speed .}%
\label{ternSlow}%
\end{figure*}

While the speed effect is taken into account for the classification Na-enhanced and Na-rich spectra (in accordance with the dependency on Fig. \ref{speedNaErr}), the distinction between normal type and Na-enhanced (Na-rich) meteors is not yet clearly defined and is particularly difficult for very slow meteors. Following the speed curve on Fig. \ref{speedNaErr} as well as similar dependencies from other relevant spectral surveys \citep{2005Icar..174...15B, 2015A&A...580A..67V, 2019A&A...621A..68V}, Na enhancement is still in most cases linked with low meteor speed. 

The ternary diagrams displaying spectral classification of meteors for different speed ranges (Fig. \ref{ternSlow}) show that various degrees of Na enhancement are already apparent for all meteors with entry speed $v_i <$ 27 km\,s\textsuperscript{-1}. Meteors with $v_i <$ 15 km\,s\textsuperscript{-1} were identified mainly as Na-rich or Fe-rich.

A reasonable assumption is that the majority of slow meteoroids will be composed of standard chondritic material, similarly to meteoroids with typically higher speeds. Spectra with Na/Mg/Fe intensity ratios close to the expected values for chondritic composition are defined as normal type \citep{2005Icar..174...15B}. In a survey of medium-sized meteoroids, normal type spectra constituted 72\% of the entire sample \citep{2019A&A...629A..71M}. Only 53\% percent of normal type spectra are observed among meteors with $v_i <$ 27 km\,s\textsuperscript{-1} and only 39\% for meteors with  $v_i <$ 20 km\,s\textsuperscript{-1}. On the other hand, while Na-rich meteors only represented 2\% of the entire sample, they constitute 26\% of meteors with $v_i <$ 20 km\,s\textsuperscript{-1} and 50\% of meteors with $v_i <$ 15 km\,s\textsuperscript{-1}. There is no physical reason for why most very slow meteoroids would be of atypical composition.

\subsection{Na-rich spectra from laboratory}

The effect of enhanced Na line intensity in slow meteors was also observed during our experimental campaign focused on simulated ablation of meteorites in high-enthalpy plasma wind tunnel, performed in cooperation with the Institute of Space Systems, University of Stuttgart. The experimental facility originally developed for re-entry modeling has been successfully tested to simulate meteor ablation \citep{2017ApJ...837..112L, 2018A&A...613A..54D}. The experiment recreates the atmospheric flight scenarios equivalent to air friction of an atmospheric entry speed of approximately 10 km\,s\textsuperscript{-1} at 80 km altitude. 

These conditions correspond to the lower limit of meteor speeds and are thus suitable for comparison with the slowest meteors in our sample. While the compared Na-rich spectra of very slow meteors ($v_i <$ 15 km\,s\textsuperscript{-1}) were typically observed at lower altitudes (70 -- 50 km), the spectra did not change significantly with altitude. More apparent changes in the meteor spectra can be observed during bright flares related to meteoroid disruption (discussed in Section \ref{nonNa}), which could not be simulated in the laboratory experiment. Three meteorite samples were tested during our first campaign: H5 chondrite Ko\v{s}ice, L3-6 chondrite Northwest Africa 869 and enstatite chondrite EL melt rock Al Haggounia 001.

\begin{figure}[t]
\centerline{\includegraphics[width=9cm,angle=0]{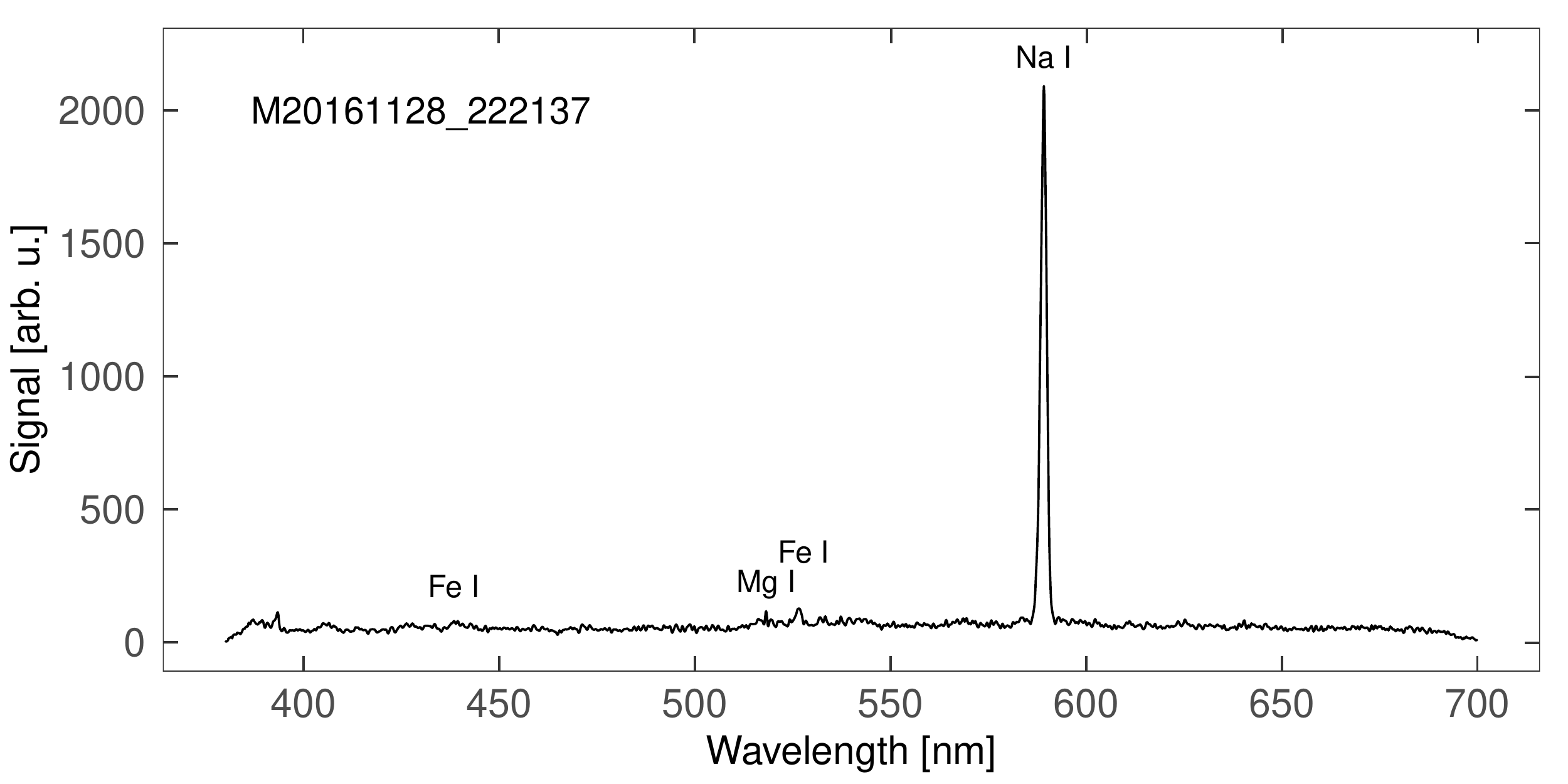}}
\centerline{\includegraphics[width=9cm,angle=0]{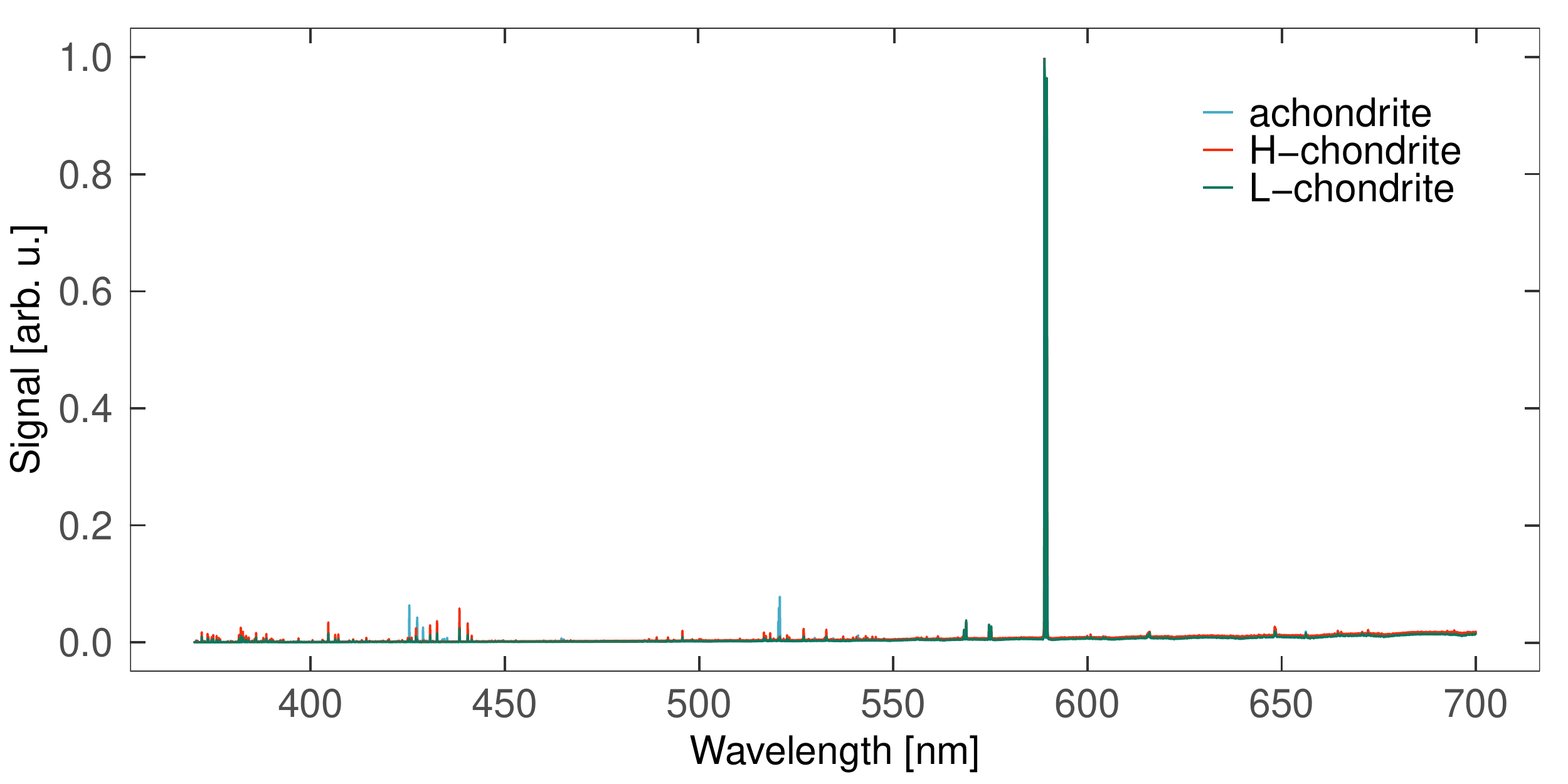}}
\caption[f1]{Spectrum of a Na-rich meteor with initial speed $v_i =$ 12.7 km\,s\textsuperscript{-1} at 83.5 km altitude observed by the AMOS-Spec-HR (upper) and Echelle spectra of three meteorite samples (lower) of different composition observed during simulated ablation in the plasma wind tunnel at conditions characteristic for flight at 10 km\,s\textsuperscript{-1} at 80 km altitude. The Echelle spectra of meteorite samples were normalized at the position of the strongest Na I line (588.99 nm).} 
\label{EchelleAll}
\end{figure}

The observed high-resolution Echelle spectra of the simulated ablation exhibited very high intensity of the Na line irrespective of the meteorite composition (Fig. \ref{EchelleAll}). The relative intensities of other lines were up to 10 times lower. The experiment reveals that the apparent Na-rich spectra are natural for meteors of very low speed at higher altitudes independently of their real composition (presumably with the exception of Na depleted bodies). Fainter spectral lines are simply lost in the background noise. The comparison of the overall spectrum with the slowest meteor in our sample M20161128\_222137 (Fig. \ref{EchelleAll}) shows a very good match with the experimental data. Based on the identified Mg I and Fe I lines in the 510-550 nm region, the meteor spectrum fits the best with chondritic composition.

Out of the three meteorite samples, the highest Na/Mg, Na/Fe intensity ratios were determined in the spectrum of EL chondrite melt rock Al Haggounia 001. This heavily weathered meteorite was originally classified as aubrite, but recently reclassified by \citet{doi:10.1111/maps.12679}. The spectrum reflects the meteorite's low iron content, but showed surprisingly strong Cr I lines in the 425-430 nm and 520-522 nm region. The high Cr I line intensities probably reflect the relatively high content of daubreelite (FeCr\textsubscript{2}S\textsubscript{4}) and the low excitation of the Cr I lines. Still, the main component of this class is enstatite (MgSi\textsubscript{3}O) while minerals containing Cr and Na only form a minor fraction. The high intensities of Na and Cr lines suggest a specific ablation and radiation conditions, probably low temperature and incomplete evaporation (Dr. J. Borovi\v{c}ka, private communication). Similar conditions probably occur during the ablation of smaller slow meteoroids and might be the cause of the anomalous spectra. The strong Cr I lines in the spectrum of Al Haggounia 001 however must reflect enhanced Cr content in comparison to the H and L chondrites, as the experiment conditions were the same for each sample. Similar spectra were not detected among meteors in our survey.

If we assume intensity ratio Na/Mg in the range 0.7 - 1.5 for a meteorite of chondritic composition (valid assumption following our results \citep{2019A&A...629A..71M} and the modeling of \citet{2005Icar..174...15B}), we can estimate that the Na line intensity is increased by a factor of 40 to 95 in the spectrum of the H-chondrite observed at simulated flight conditions of 10 km\,s\textsuperscript{-1}. 

\begin{figure}[t]
\centerline{\includegraphics[width=8.5cm,angle=0]{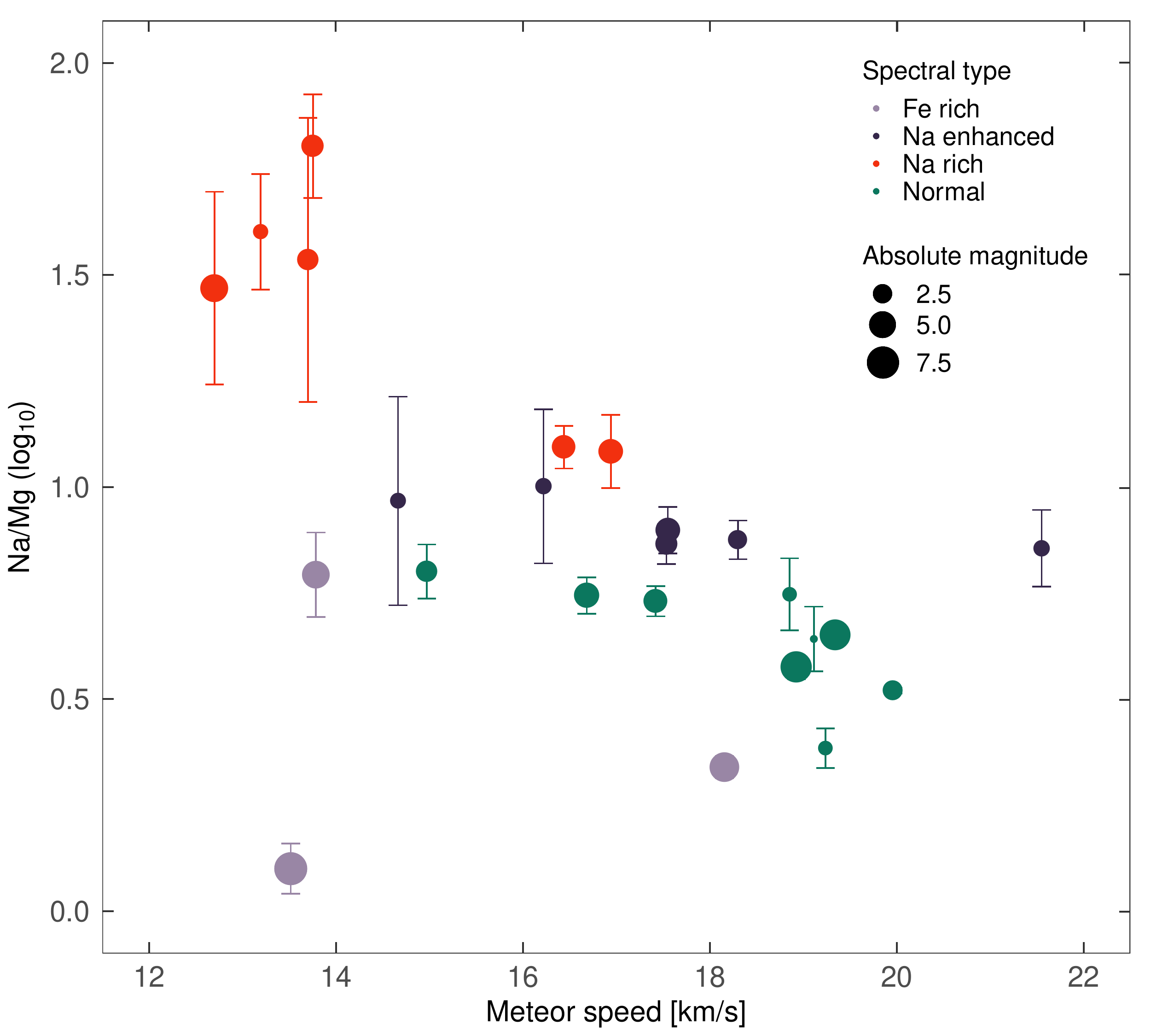}} \caption[f1]{Zoomed in section of Fig. \ref{speedNaErr} for meteors with $v_i <$ 22 km\,s\textsuperscript{-1}. Meteoroid marker sizes reflect relative meteor magnitudes. Meteors brighter than -5 mag were not classified as Na-rich or Na-enhanced, with the exception of the slowest meteor in our sample.} 
\label{speedNazoom}
\end{figure}

It should be noted that while we were able to identify the main Mg I and Fe I lines in the spectrum of Na-rich meteor M20161128\_222137 (Fig. \ref{EchelleAll}), they are often not detected at all in Na-rich meteor spectra. We believe that this is simply caused by the low S/N of these lines, related to meteor brightness. Indeed, most Na-rich meteors have moderate magnitudes. The brightest low-velocity meteors, especially with visible flares, have Fe-rich or normal-type spectra (Fig. \ref{speedNazoom}). No meteors brighter than -5 mag were classified as Na-rich or Na-enhanced, with the exception of the slowest meteor in our sample M20161128\_222137, which however showed lower Na/Mg ratio compared to the expected trend on Fig. \ref{speedNazoom}.

Based on the presented arguments, it is probable that most previously identified slow Na-enhanced and Na-rich meteors actually represent normal type (chondritic) bodies with increased Na intensity reflecting the specific ablation conditions. In that case, the dependency of mean Na/Mg intensity ratio on meteor speed would need to be steeper near the lower limit meteor speeds (dashed line on Fig. \ref{speedNaErr}). The slope of this dependence is however affected by the uncertainty of the positions of Na-rich meteors on this plot, related to the very low intensity of the compared Mg line. More accurate description of the speed dependency could be achieved by obtaining spectra of a meteorite of the same composition at different flight conditions (higher speed and lower altitudes). 

It is possible that some Na-enhanced and Na-rich meteors identified by \citet{2005Icar..174...15B}, \citet{2015A&A...580A..67V} and \citet{2019A&A...621A..68V} among mm-sized meteoroids may also be to some degree still affected by the low meteor speed. The correction for meteor speed might however not be completely straightforward. There are reports of slow Na-rich meteors which could represent meteoroids of unusual composition. Such estimate can be made if other atypical emission lines are identified in the meteor spectrum. One such example is meteor SZ 1440 ($v_i \approx$ 26 km\,s\textsuperscript{-1}) detected by \citet{2005Icar..174...15B} with strong Na lines and several lines of the refractive Ca. Another unusual meteor with very low speed ($v_i =$ 10.9 km\,s\textsuperscript{-1}) described by \citet{2016AJ....151..135C} showed spectrum with dominant Na and relatively strong Fe lines. Another Na-enhanced meteors ($v_i =$ 19.6 km\,s\textsuperscript{-1}) with possibly atypical composition was described by \citet{2008EM&P..102..485B}.

Spectra with strong Na line and no other measurable lines remain opened to interpretation. We report that they can be explained by normal composition and the effect of meteor speed and moderate brightness, and this explanation seems to be statistically the most probable. 

\subsection{Slow meteors not dominated by Na} \label{nonNa}

Besides the Na-rich and Na-enhanced meteors, the only distinct spectral type detected among slow meteors ($v_i <$ 22 km\,s\textsuperscript{-1}) is the Fe-rich (Fig. \ref{ternSlow}). Generally, Fe-rich meteors are caused by bodies with real enhanced Fe content associated with Fe-rich H-type chondrites or possibly stony-iron meteoroids \citep{2019A&A...629A..71M}. Light curves of slow Fe-rich meteors differ from the Na-dominated meteors by exhibiting one or multiple bright flares associated with sudden disruption of the meteoroid. 

\begin{figure}[!t]
\centerline{\includegraphics[width=9cm,angle=0]{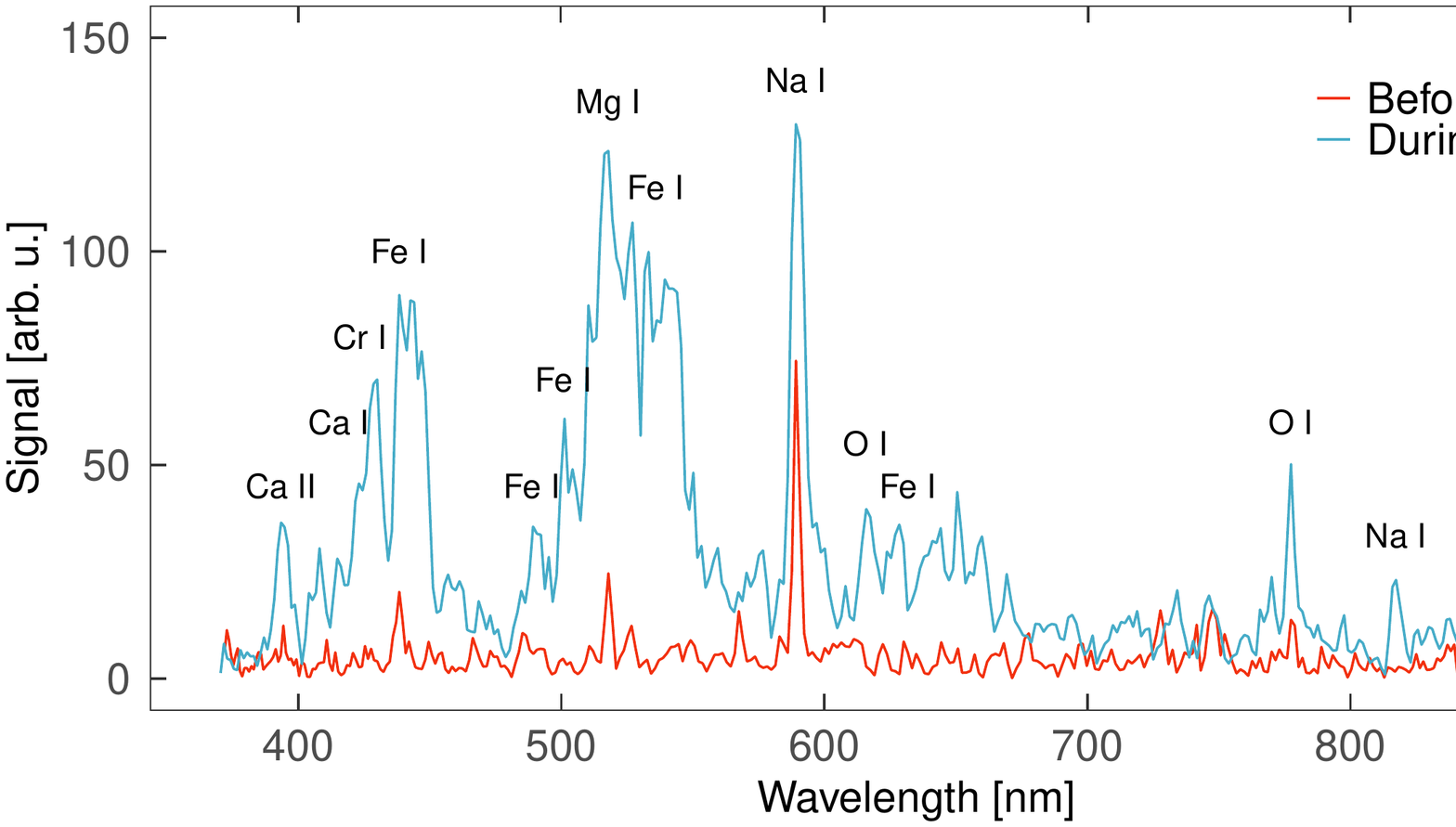}}
\centerline{\includegraphics[width=9cm,angle=0]{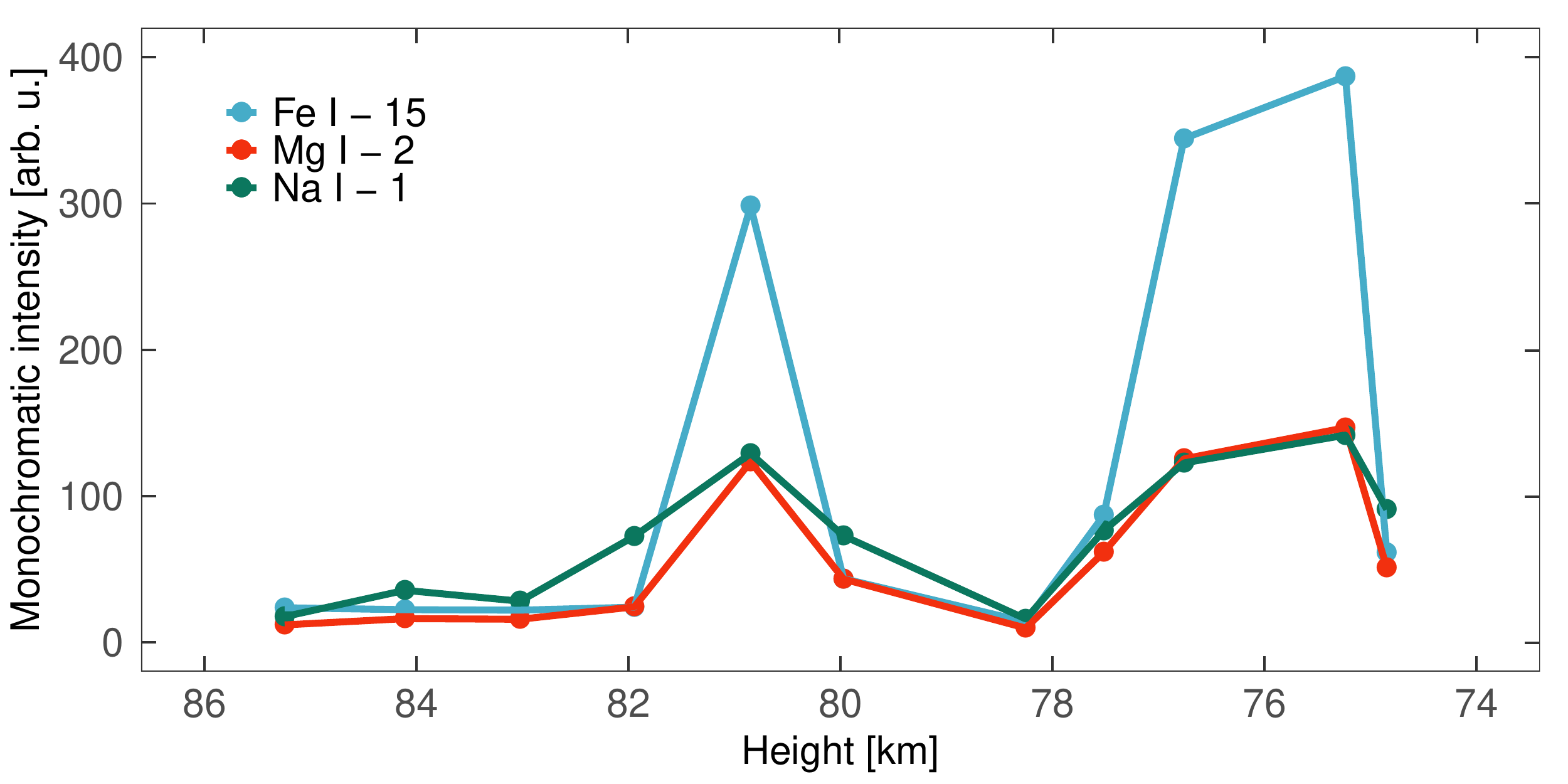}} \caption{Upper: comparison of a spectrum observed during one frame shortly before the flare (red) and during the flare (blue) of meteor M20160727\_000019. The displayed profile is not corrected for the spectral sensitivity of the system (Fig. \ref{Sens}). Lower: monochromatic light curves of this meteor showing apparent increase of Fe intensity during the flare.} 
\label{flare_example}
\end{figure}
 
One such example is meteor M20160727\_000019, which was classified as borderline normal type, but shows representative behavior similar to slow Fe-rich meteors. The comparison of the spectrum shortly before the first bright flare and during the flare can be seen on Fig. \ref{flare_example}. It is apparent that emission from most of the lines originated in the flare, while Na was dominant before the disruption event. Similarly, monochromatic light curve at Fig. \ref{flare_example} shows a sharp increase of Fe intensity during the two exhibited flares, while Na line remained dominant before and after the flare. The effects of saturation and self-absorption caused by optically thick plasma can affect the observed meteor spectra. The apparent significant increase of Fe/Na and Fe/Mg intensity ratio observed during the flare (Fig. \ref{flare_example}) is partially caused by this effect. Similar frames cannot be used for the line intensity ratio measurements, but can reveal fainter emission lines which would remain undetected in moderate Na-enhanced spectra.

We noted that the detected dependency of Na/Mg ratio on meteor speed relates to the achieved temperature, which is probably the main factor causing the observed Na enhancement. During the flares, we do not assume significant rise of temperature, but rather increase of brightness related to disruption and consequent expanse of the cross-section of the meteor producing the emission. Based on our experiences, in meteors of moderate magnitudes ($<$ -5 mag) and entry speeds ($<$ 22 km\,s\textsuperscript{-1}), the low-excitation Na line is preferred from the start of the ablation, while other lines (mainly Mg and Fe) radiate at intensities near the noise level of the recording.

\section{New classification and confirmed Na-enhanced and Na-rich meteoroids} \label{Reclass}

\begin{figure}
\centerline{\includegraphics[width=8.5cm,angle=0]{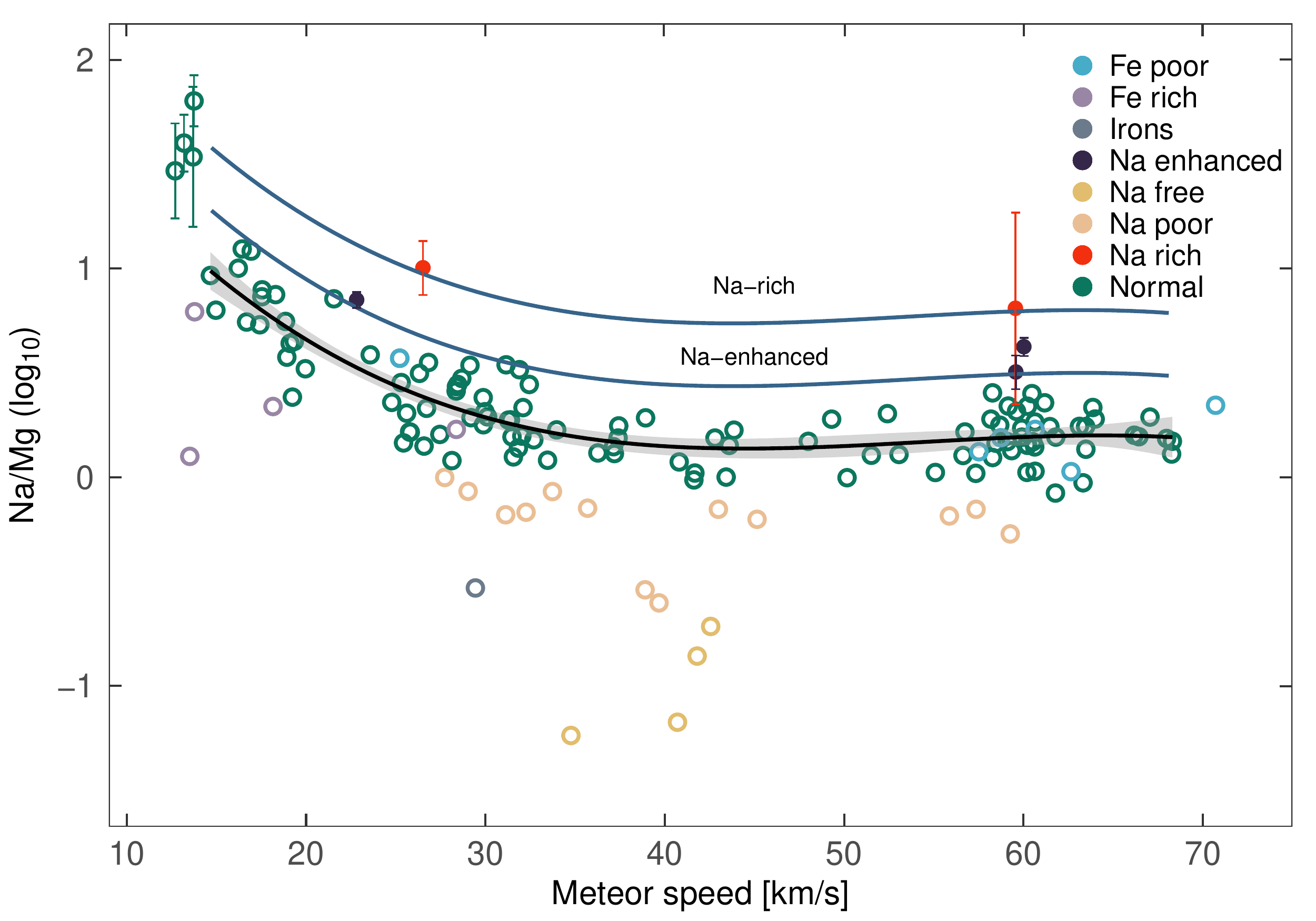}} \caption[f1]{The proposed boundaries for the classification of Na-enhanced and Na-rich meteors displayed on the same data as in Fig. \ref{speedNaErr}. More spectral measurements for meteors slower than 15 km\,s\textsuperscript{-1} are necessary to establish boundaries in this region.} 
\label{ReClas}
\end{figure}

Given the disparity in the classification of Na-enhanced and Na-rich meteoroids, we propose new speed-dependent boundaries of these spectral classes (Fig. \ref{ReClas}). The regions are defined by lines corresponding to an increase of the moving average Na/Mg intensity ratio in normal type meteors by a factor of Na/Mg = 2. This means that the boundary of a Na-enhanced class is positioned for any given speed at (Na/Mg)\textsubscript{avr} x 2 and the boundary of a Na-rich class at (Na/Mg)\textsubscript{avr} x 4. 

The corresponding equation for the fit of the average Na/Mg intensity ratio for normal type meteors is based on our data: (Na/Mg)\textsubscript{avr} = 2.4994 - 0.138\,$v_i$ + 0.0026\,${v_i}^2$ - 0.00002\,${v_i}^3$. It is however recommended that the classification of Na-enhanced and Na-rich meteors is made relative to the fit of normal type meteors observed by the same instrument, as the average Na/Mg can depend on meteoroid size \citep{2019A&A...629A..71M} and therefore also on the instrument sensitivity.

More spectral data for very slow meteors are needed to properly define boundaries for $v_i <$ 15 km\,\textsuperscript{-1}. We note that the fit of the running average of the Na/Mg intensity ratio for normal type meteors (Fig. \ref{ReClas}) was fit by a 3rd degree polynomial, but may not follow this function for $v_i <$ 15 km\,\textsuperscript{-1}. The results of this work suggest that most very slow Na-rich meteors probably represent common chondritic bodies and could therefore be classified as normal. However, before the boundaries in this region are established, very slow meteors with apparently high Na/Mg intensity ratio can be considered as Na-rich/Na-enhanced candidates and should be evaluated individually also considering their orbital and physical data.

Following the proposed classification (Fig. \ref{ReClas}), three of the studied meteoroids in this work can be classified as Na-enhanced (one Perseid, one $\nu$-Draconid and one sporadic meteor), and two are on the boundary but classify as Na-rich (one Perseid and one $\alpha$-Capricornid). Based on the orbital and material properties discussed in Section \ref{candidates}, all five meteoroids with confirmed enhancement of Na are of cometary origin.

Three of these meteors were found with speed close to 60 km\,s\textsuperscript{-1} (Table \ref{strNaE}). Here, the compositional Na enhancement is easily revealed as it is not affected by the speed dependency (Fig. \ref{ReClas}). One fast Na-enhanced meteor was also observed by \citet{2019A&A...621A..68V}.

\begin{figure}[t]
\centerline{\includegraphics[width=9cm,angle=0]{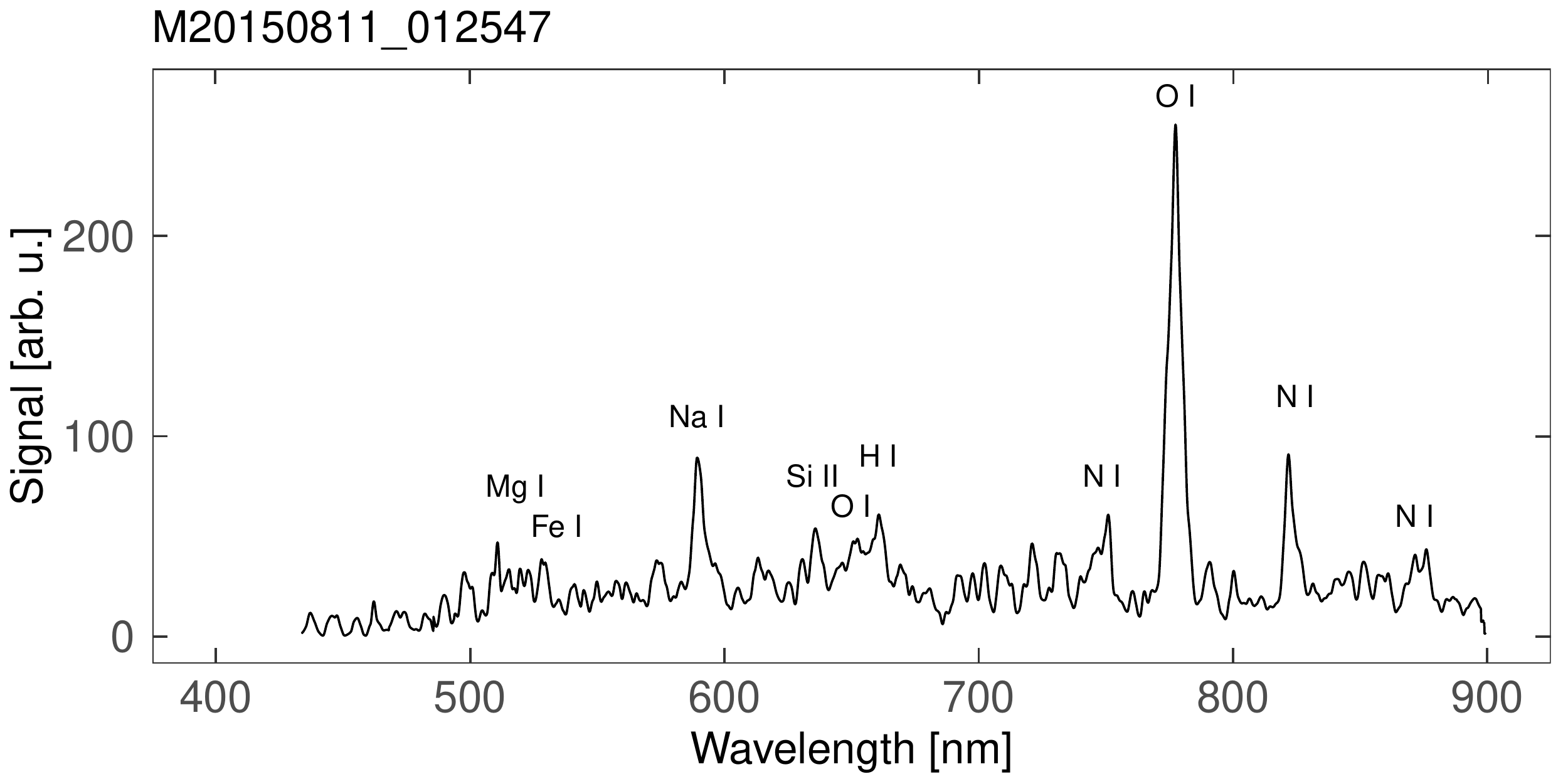}}
\centerline{\includegraphics[width=9cm,angle=0]{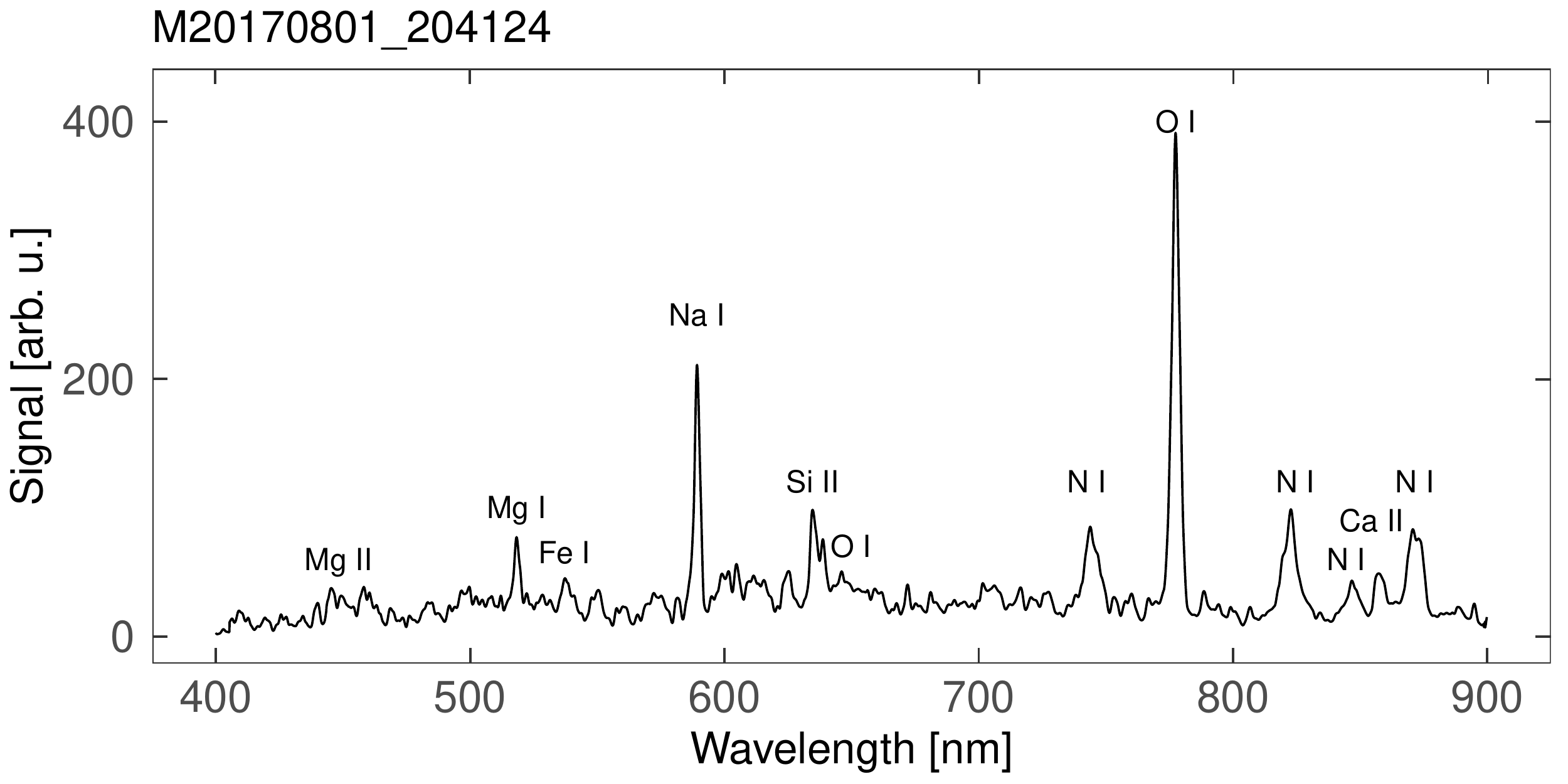}} \caption[f1]{Spectra of two Perseids confirmed as Na-enhanced and Na-rich. The displayed profiles are not corrected for the spectral sensitivity of the system (Fig. \ref{Sens}).} 
\label{NaE_fast}
\end{figure}

Two of the meteors with confirmed Na enhancement are members of the Perseid stream (Fig. \ref{NaE_fast}), which originates in comet 109P/Swift-Tuttle. The volatile-rich cometary bodies could be the natural source of Na-enhanced composition. Yet, most of the cometary meteors and specifically Perseids are classified as normal type or Fe-poor \citep{2019A&A...629A..71M}. The detected Na enhancement could reflect natural heterogeneity of the volatile content among cometary bodies or could relate dynamically young debris with bulk composition unaltered by previous space weathering effects.

Surprisingly, we determined relatively high material strength (type A/II and type C1/II) of these two meteors (Table \ref{strNaE}), characteristic for carbonaceous materials or denser cometary bodies. By comparison with the properties of 19 Perseids analyzed by \citet{2019A&A...629A..71M}, such high material strength is not characteristic for most Perseids, although it was also detected in two samples with normal type spectra. The increased material strength might be linked with the detected Na enhancement. The study of \citet{2019A&A...621A..68V} pointed out that the increased Na content is often associated with meteoroids consisting of larger grains.  

Besides Perseids, two more meteors associated with a cometary meteoroid stream were found with increased Na content. A $\nu$-Draconid (IAU MDC \#220), originating from a minor stream with unknown parent object and a $\alpha$-Capricornid (IAU MDC \#001), which originates in comet 169P/NEAT \citep{2010AJ....139.1822J}.

Overall four $\alpha$-Capricornid spectra were observed in our previous survey \citep{2019A&A...629A..71M}, including two meteors exhibiting Na enhancement (classified as Fe-poor and Na-enhanced). The remaining two meteors displayed behavior represented by meteor M20160727\_000019 (Fig. \ref{flare_example}), which was defined by two major disruption events accompanied by a bright flare with emission of numerous lines (Fig. \ref{flare_example}). In contrast, Na-enhanced $\alpha$-Capricornids had smooth light curves without apparent explosions of brightness. 

Assuming that the parent comet 169P/NEAT is of relatively homogeneous composition, the significant variations of the detected spectra are likely caused by the low brightness of non-Na lines in meteors exhibiting moderate magnitudes and no flares on one side, and the change of the optical thickness during bright flares on the other. The observed distinct light curves of the $\alpha$-Capricornids more likely reflect their structural heterogeneity and significant porosity. Overall material strength of all studied $\alpha$-Capricornids was low, typical for cometary bodies (type C1/IIIA/IIIB). The observed diversity might also present higher volatile content in $\alpha$-Capricornids, which is released gradually or suddenly during flares. 

\section{Discussion}

\subsection{Possibilities for real enhancement of Na}

Based on the arguments presented in this work, we assume that most of the investigated Na-rich meteors originate in meteoroids composed of common chondritic material. Nevertheless, there are cases of slow Na-rich spectra \citep[e.g.][]{2005Icar..174...15B} which could be caused by meteoroids of atypical composition. In the following discussion, we provide overview of the identified candidates for atypical interplanetary materials which could fit the observed properties of Na-rich meteors. We consider meteors with enhanced content of Na and indications of notable refractory (particularly Ca) content. The determined short-period heliocentric orbits and increased material strengths were taken into account. From our review, we have selected three possible interpretations of Na-rich meteors assuming atypical composition:

\begin{enumerate}

\item
Composition rich in low-anorthosite plagioclase minerals, suggesting relation to ejected martian or possibly lunar material \citep{papike1998planetary,norton2008field}.

\vspace{0.3cm}

Plagioclase minerals can be found in meteorite samples in different forms ranging from anorthites (Ca-rich) to albites (Na-rich). While quite rare, minor amounts of albite can be found in the SNC meteorites, which in general exhibit enhanced content of plagioclase minerals \citep{papike1998planetary, norton2008field}. Moreover, shock during the ejection of specimens from Mars (or by prior impact) can form diaplectic glassy form of plagioclase feldspar called maskelynite \citep{tschermak1872meteoriten}. With sodium being a significant spectral feature even in meteoroids with chondritic composition, it is possible that Na-rich spectra could reflect only moderate enhancement in Na abundance. This explanation also provides reasonable fit when accounting for the rough structural and orbital properties of Na-rich meteors. Considering the determined densities and porosities of SNC meteorites \citep{1998M&PS...33.1231C, 2007M&PS...42.2043C}, they are in terms of strength expected to lay between carbonaceous and ordinary chondrites, which is in agreement with our data. Orbits of some of the observed Na-rich meteors with $1.2 < a < 1.8 $ (Table~\ref{orbNaR}) also do not rule out that these bodies evolved from Mars and drifted to more eccentric orbits under the influence of gravitational interactions with larger bodies. Based on the measured cosmic-ray exposure ages, SNC samples are among the youngest known meteorites \citep{1997GeCoA..61.2749E}. The efficiency of delivery of martian ejecta to Earth is up to 7.5\%, with about a third of the encounters with Earth occurring within 10 Myr \citep{1997Icar..130..228G, papike1998planetary}.

Assuming similar mineralogical content, another possible interpretation of the origin of Na-rich meteoroids could be ejected lunar feldspathic rock. Although lunar rocks are typically low in sodium, the varied lunar geology could provide low-anorthosite material that could explain the observed spectra. In this case, longer orbital evolution affected by close approaches to Sun and Earth would be required to fit the observed heliocentric orbits.

\vspace{0.3cm}

\item
Composition related to differentiated C-type or D-type asteroids. This follows the recent discovery of large amounts sodium carbonates on Ceres \citep{2015Natur.528..241D}.

\vspace{0.3cm}

Recent discovery of salts in bright spots of Ceres, formed by crystallization of brines that reached the surface raised new view on the processes by which materials from outer solar system, including organic matter, can be incorporated into asteroidal bodies during early stages of the solar system formation \citep{2015Natur.528..241D, 2015Natur.528..237N}. These materials include large amounts of sodium carbonates (Na\textsubscript{2}CO\textsubscript{3}), and smaller amounts ammoniated phyllosilicates NH\textsubscript{4}Cl or NH\textsubscript{4}HCO\textsubscript{3} \citep{2016Natur.536...54D, 2017P&SS..141...73V}. Considering the stable, low-eccentric orbit of Ceres with no dynamical family or associated meteorite impacts, it is unlikely that it would produce significant number of smaller Na-rich meteoroids. It is however believed that at least some other large low-albedo asteroids could have similar geologic histories (Dr. A. Rivkin 2018, private communication). While it is unclear if such content can be effectively preserved in cm-m sized bodies, this discovery raises the possibility that meteoroids with real Na-rich content could be related to differentiated asteroids, which have previously accreted material from beyond the 'snow line'.

\vspace{0.3cm}

\item
The presence of Na-rich chondrules, which can occur mostly in ordinary and carbonaceous chondrites as part of Al-rich objects \citep{2016GeCoA.177..182E}.

\vspace{0.3cm}

Na-rich chondrules occur mostly in ordinary and carbonaceous chondrites as part of Al-rich objects (Ca,Al-rich inclusions, Al-rich chondrules, Al-rich fragments) \citep{1984GeCoA..48..693B}. These objects with significant concentrations of Na\textsubscript{2}O were formed by melting of precursors containing an (ultra-)refractory element-rich component and Na-rich constituents \citep{2016GeCoA.177..182E}. It is not clear if small debris rich in Na-rich chondrules could produce considerable meteoroid population.

\end{enumerate}

\subsection{Future aims}

Given the strong influence of the speed dependence on the observed Na-rich spectra, composition of these bodies cannot be confirmed from low-resolution video spectra in standard sensitivity. Further study would require detailed analysis of higher-resolution spectra captured by very sensitive cameras. Inspection based on detailed dynamical integration assuming meteoroid ejection from larger body could provide confirmation of the found potential parent Apollo asteroids.

We intend to further study the effect of meteor speed on the observed Na/Mg intensity ratio by observing emission spectra of laboratory simulated ablation of various meteorite samples at different physical conditions. Our first results confirm the observations of Na-rich spectra for meteorites of various composition at conditions corresponding to atmospheric flight of approximately 10 km\,s\textsuperscript{-1} at 80 km height. From the observed speed curve (Fig. \ref{ReClas}), it is clear that the effect is present for all meteors slower than 40 km\,s\textsuperscript{-1}.

Recent work by \citet{2016AJ....151..135C} reported on a first impact detection of a temporarily captured natural satellite. The original meteoroid showed properties similar to the analyzed Na-rich meteors with very low meteor speed (11.0 km\,s\textsuperscript{-1}) and spectrum dominated by Na and Fe. Meteoroids with Na-rich spectra would therefore be good candidates for investigating previous temporary captures. For confident confirmation, such investigation would require very high precision of the determined entry velocity.

The composition of very slow Na-rich meteors remains open to interpretation. The confirmation of volatile-rich asteroidal fragments would be of high scientific interest. It was suggested that volatile elements including water could have been brought to Earth by meteorites formed in the inner solar system during the first two million years \citep{2017RSPTA.37560209S, 2017GeCoA.212..156S}.

\section{Conclusions}

We have presented the first in-depth analysis of Na-enhanced and Na-rich meteors. First observed by \citet{2005Icar..174...15B}, these spectral classes were defined to designate meteoroids which deviate from the expected chondritic composition at given speed. Based on our results, the majority of identified Na-enhanced and Na-rich spectra can be explained by the effect of low meteor speed related to low ablation temperatures and common chondritic composition. The only other distinct spectral group observed among slow meteors are the Fe-rich, which are typically linked with bright flares, unlike the typically moderately bright Na dominated meteors.

Laboratory experiment using simulated ablation of known meteorite samples at conditions corresponding to atmospheric flight of a very slow meteoroid (approximately 10 km\,s\textsuperscript{-1} at 80 km height) has shown that apparent Na-rich spectra are observed irrespectively of the meteorite composition. The obtained spectra show that at such conditions and considering the typical S/N of the meteor spectra observations, other major lines would often be lost in the background noise, as is observed in most Na-rich meteors. We estimate that for an H-type chondrite with speed of $\approx$ 10 km\,s\textsuperscript{-1}, the Na line intensity is increased by a factor of 40 to 95.

Based on the determined heliocentric orbits and material strengths, we can conclude that all previously identified Na-rich meteors represented bodies of stronger material comparable to carbonaceous or ordinary chondrites and originate from Apollo-type asteroidal orbits. We therefore assume that most of these meteoroids are likely composed of chondritic material and represent fragments of Apollo asteroids.

For more consistent classification of Na-enhanced and Na-rich \textit{meteoroids} in the future, we propose new speed-dependent boundaries of these spectral classes (Fig. \ref{ReClas}), defined by an increase from the moving average Na/Mg intensity ratio in normal type meteors at any given speed. Due to a small data sample, the boundaries are not yet defined for meteors slower than 15 km\,\textsuperscript{-1}. Very slow meteors with spectra dominated by Na should be interpreted individually. Our work reports that most Na-rich and Na-enhanced meteors linked with low speed can be explained by the low ablation temperature and common chondritic composition.

Real compositional Na enhancement in meteoroids can be unambiguously revealed among faster cometary meteors. We confirm Na enhancement in five cometary meteoroids (three Na-enhanced and two Na-rich): two Perseids, an $\alpha$-Capricornid, a $\nu$-Draconids and a sporadic. Both Na-enhanced Perseids were linked with increased material strengths compared to other stream members, suggesting that the detected increase of volatile content has implications for the meteoroid structure. This effect could relate the previously reported larger grain sizes in meteoroids with increased Na content \citep{2019A&A...621A..68V}. 

\section*{Acknowledgements}

We are thankful to J. Borovi\v{c}ka and the anonymous referee for their helpful reviews, which improved this paper. This work was supported by the Slovak Research and Development Agency grant APVV-16-0148 and the Slovak Grant Agency for Science grant VEGA 01/0596/18. We are grateful to the support staff contributing to the operation of the AMOS network in Slovakia and to the Instituto de Astrof\'{i}sica de Canarias for providing support with the installation and maintenance of AMOS systems on the Canary Islands. The AMOS detection program was developed in cooperation with the KVANT company.

\bibliography{references}

\end{document}